\documentclass[sigconf]{acmart}
\usepackage{xcolor}
\usepackage{subcaption}
\usepackage{hyperref}
\usepackage{array}
\usepackage{multirow}
\AtBeginDocument{%
  \providecommand\BibTeX{{%
    \normalfont B\kern-0.5em{\scshape i\kern-0.25em b}\kern-0.8em\TeX}}}

  {\list{}{\leftmargin=0.2in\rightmargin=0.2in}\item[]}%
  {\endlist}

\copyrightyear{2024}
\acmYear{2024}
\setcopyright{rightsretained}
\acmConference[CHI '24]{Proceedings of the CHI Conference on Human Factors in Computing Systems}{May 11--16, 2024}{Honolulu, HI, USA}
\acmBooktitle{Proceedings of the CHI Conference on Human Factors in Computing Systems (CHI '24), May 11--16, 2024, Honolulu, HI, USA}
\acmDOI{10.1145/3613904.3642738}
\acmISBN{979-8-4007-0330-0/24/05}

\begin{document}

\title[Designing for Community Stakeholders' Interactions with AI in Policing]{Are We Asking the Right Questions?: Designing for Community Stakeholders' Interactions with AI in Policing}

\author{MD Romael Haque}
\authornote{Both authors contributed equally to this research.}
\affiliation{%
  \institution{Marquette University}
  \city{Milwaukee}
  \state{WI}
  \postcode{53233}
  \country{USA}
}
\email{mdromael.haque@marquette.edu}

\author{Devansh Saxena}
\authornotemark[1]
\affiliation{%
  \institution{Carnegie Mellon University}
  \city{Pittsburgh}
  \state{PA}
  \postcode{15289}
  \country{USA}
}
\email{devanshsaxena@cmu.edu}

\author{Katy Weathington}
\affiliation{%
  \institution{University of Colorado Boulder}
  \city{Boulder}
  \state{CO}
  \postcode{80309}
  \country{USA}
}
\email{katy.weathington@colorado.edu}

\author{Joseph Chudzik}
\affiliation{%
  \institution{University of Chicago}
  \city{Chicago}
  \state{IL}
  \postcode{60637}
  \country{USA}
}
\email{jchudzik@uchicago.edu}

\author{Shion Guha}
\affiliation{%
  \institution{University of Toronto}
  \city{Toronto}
  \state{ON}
  \country{Canada}
}
\email{shion.guha@utoronto.ca}

\renewcommand{\shortauthors}{Haque and Saxena et al.}


\begin{abstract}
Research into recidivism risk prediction in the criminal justice system has garnered significant attention from HCI, critical algorithm studies, and the emerging field of human-AI decision-making. This study focuses on algorithmic crime mapping, a prevalent yet underexplored form of algorithmic decision support (ADS) in this context. We conducted experiments and follow-up interviews with 60 participants, including community members, technical experts, and law enforcement agents (LEAs), to explore how lived experiences, technical knowledge, and domain expertise shape interactions with the ADS, impacting human-AI decision-making. Surprisingly, we found that domain experts (LEAs) often exhibited anchoring bias, readily accepting and engaging with the first crime map presented to them. Conversely, community members and technical experts were more inclined to engage with the tool, adjust controls, and generate different maps. Our findings highlight that all three stakeholders were able to provide critical feedback regarding AI design and use - community members questioned the core motivation of the tool, technical experts drew attention to the elastic nature of data science practice, and LEAs suggested redesign pathways such that the tool could complement their domain expertise.
\end{abstract}



\maketitle

\section{Introduction}

Predictive policing stands as a significant and contentiously debated topic within the criminal justice system (CJS) today \cite{perry2013predictive}. To pre-emptively identify where crime is likely to occur and who is likely to commit it (i.e., preventing crime before it happens), police agencies have begun to supplement conventional forecasting and patrolling practices with the use of algorithmic decision-support systems (ADS) \cite{simmons2016quantifying}. One popular approach, called algorithmic crime mapping \cite{haque2020understanding} involves using geographic data to identify hotspots where crime is expected to occur in the future to assist LEAs (Law Enforcement Agents) in proactively and efficiently allocating resources \cite{perry2013predictive}. Moving to data-driven policing, however, introduces a variety of concerns \cite{borradaile2020whose}. These systems rely on a large amount of local and geospatial data but fail to account for the trustworthiness or credibility of these different sources of data \cite{clancy2022reconciling}. Moreover, it is critical to understand the association between crime analysis and how it justifies resource allocation because it is often hard for LEAs to determine how to best incorporate empirical insights within existing work practices in order to improve decision-making \cite{haque2019exploring}. Moreover, the introduction of ADS in these complex real-world contexts is expected to remove human biases from the decision-making process and make it more objective. However, growing evidence suggests that these systems exacerbate human biases embedded in the historical data \cite{browning2021stop, lum2016predict, veale2018fairness} as well as introduce new biases as a result of the inconsistency between the workers' decision-making latitude versus the empirical decisions recommended by the ADS \cite{pruss2023ghosting, saxena2021framework, cheng2022child}. These issues have underscored the calls for developing human-centered systems \cite{jarrahi2023principles, aragon2022human, saxena2020human} that situate empirical predictions in the context of work and complement workers’ expertise \cite{kawakami2023training, saxena2023algorithmic}, i.e. - allow stakeholders such as law enforcement and crime analysts to derive meaningful information that aligns with their work and informs decision-making on interventions and resource allocation. 

However, to ensure that ADS are human-centered, particularly in the public sector, it's crucial to integrate the perspectives and needs of the stakeholders who are most impacted by the use of such systems, i.e. - community members. Here, SIGCHI researchers have begun collaborating directly with community members affected by ADS \cite{kuo2023understanding, stapleton2022has, zhang2023deliberating, nielsen2023cares}, in addition to civil servants whose work is impacted by such technology \cite{saxena2023algorithmic, kawakami2023training, ammitzboll2021street}. Engaging with a diverse range of community stakeholders enables us to gain a deeper understanding of what it means for ADS used in government agencies to operate in the public interest and further helps us identify ways to enhance the practices of civil servants who are entrusted with serving the community \cite{stapleton2022has, wan2023community}. We contribute to this body of work by introducing an intuitive AI application that community members, technical experts, and law enforcement agents can interact with. The application anchors empirical findings within city neighborhoods familiar to the participants. It served as a boundary object \cite{star1989structure} that facilitated detailed and nuanced discussions with them about policing practices in their city and the use of ADS by police departments. Furthermore, it aided in eliciting participants' feedback on the ethical design and use of such systems.

Through a human-centered lens, we examine how participants from diverse backgrounds interact with an intuitive AI application, how it informs their perspectives and needs regarding the ethical design and use of such systems, as well as uncover implications for human-AI decision-making in a public sector setting. In this study, we ask the following research questions -- 

\begin{itemize}
    \item \textbf{RQ1:} How do lived experiences, technical skills, and domain expertise shape people's perspectives of algorithmic decision-support systems (ADS) such as algorithmic crime mapping used in the public sector?
    \item \textbf{RQ2:} How do these different backgrounds translate into people's utilization of the ADS system? 
    \item \textbf{RQ3:} What are the needs of different community stakeholders regarding the ethical design and use of ADS such as algorithmic crime mapping?
\end{itemize}

We conducted a mixed-methods study in a mid-sized Midwestern U.S. city, where we developed an interactive crime-mapping application that facilitated discussions with 60 participants, including community members (n=39), technical experts (n=14), and law enforcement agents (n=7). We developed this application based on prior literature on crime-mapping techniques  \cite{chainey_utility_2008,johansson2015crime,rosser2017predictive} where it displays a heatmap resulting from a Kernel Density Estimation (KDE) algorithm \cite{kalinic2018kernel} given crime data and specific user inputs. There were two aspects to this study - 1) in a lab activity, participants set alternative parameters for the KDE algorithm and located hotspots on the map, allowing us to assess their interaction and interpretation of algorithmic crime mapping, and 2) through semi-structured interviews conducted immediately after the activity, we gained insights into participants' perspectives and concerns regarding the use of such tools by police departments as well as their needs that would ensure the ethical design and use of these tools such that they serve the public's interest. In this study, we make the following contributions -- 
\begin{itemize}

     \item We show that community stakeholders are able to provide critical feedback on various AI life stages: community members shared that the core motivation (i.e., AI problem formulation) was misaligned with the public's interest, technical experts highlighted the elastic nature of data science practice (i.e., AI development process), and LEAs suggested redesign pathways to align AI with practice and complement their domain expertise (i.e., AI interaction and use).

    \item We show that intuitive ADS, like crime mapping, can teach community members about the utility, but more importantly, the limitations of AI. They learned that the system only generated estimates that were subject to change based on their choice of parameters. They were able to contextualize findings based on their local knowledge of the city and deliberate on the significant impact of the underlying data.

    \item We found that a target-construct mismatch \cite{kawakami2023training} exists even in algorithmic crime mapping, a widely used and intuitive ADS. Law enforcement agents (LEAs) focus on different city landmarks (i.e., the construct) and not the predicted local hotspots (i.e., the target). They prioritize district boundaries and geospatial crime relationships, where the local hotspot only provided a high-level `satisfactory outcome'.

    \item Consequently, we learned that an ADS that is not worker-centered and failed to consider key attributes valued by LEAs (e.g., relationship between crimes, district boundaries) that would augment their practice further complicated human-AI decision-making and made it hard to assess the efficacy of human+AI decisions over human-only decisions.

\end{itemize}

In the following sections, we first discuss related literature on AI in policing and algorithmic crime mapping, involving stakeholders in AI deliberation and design, and human-AI decision-making that have informed this study.

\section{Related Work}
In this section, we first discuss recent research conducted on ADS use in policing followed by research conducted on algorithmic decision-making in the public sector and engaging community members in the AI deliberation process.

\subsection{AI in Policing and Algorithmic Crime Mapping}

Concerning the criminal justice system (CJS), researchers are specifically focused on unraveling stakeholders' perceptions regarding a myriad of algorithmic systems. This includes understanding judges' interpretations of diverse risk assessment tools \cite{pruss2023ghosting, stevenson2022algorithmic}, law enforcement's insights into predictive crime analysis algorithms \cite{veale2018fairness, haque2019exploring}, and similar considerations \cite{grgic2018human, grgic2016case,srivastava2019mathematical,1745-9125.2011.00240}. A rising number of studies are further attempting to understand how risk assessment algorithms affect different criminal justice outcomes such as pre-trial release, recidivism rates, etc. \cite{green2019principles,grgic2019human,vaccaro2019effects,green2019disparate}. 

Algorithmic crime mapping is the application of contemporary information processing technologies that merge geographic information system (GIS) data, digital maps, and crime data to gain deeper insights into the propagation of criminal activity \cite{haque2020understanding}. It allows law enforcement organizations to examine and correlate data sources to provide a detailed snapshot of crime episodes and related characteristics within a community or geographic area. \cite{R,book12}. It has already been applied to different types of crime, including drug incidents \cite{12}, environmental crimes \cite{chainey2013gis}, burglary \cite{chainey2013gis}, gang violence \cite{kennedy_known_1997}, burglary repeat victimization \cite{10.2307/23638645}, residential burglaries \cite{noauthor_gis_nodate}, and serial robberies \cite{Hill2002OperationalizingGT}, among others. 

Among various methodologies and tools, spatial clustering and spatiotemporal correlations, spatial ellipses, thematic mapping of geographical regions, grid thematic mapping, and kernel density estimation (KDE) are some more well-known approaches  \cite{chainey_utility_2008,chainey2013gis,1193063, Toole:2011:SCC:1989734.1989742}. KDE, for example, has lately acquired prominence in terms of practical application and has proven to be quite accurate and predictive \cite{chainey_utility_2008}. The widely used KDE method involves non-parametric estimation of crime probability density, considering parameters like grid cell size, interpolation methods, and bandwidth for precise hotspot identification \cite{boppuru2020spatio}. The user's crucial role involves setting the bandwidth, where a large value may lead to information loss, while a small one emphasizes local data \cite{johansson2015crime}. This analytical technique, applied to various crime types, transforms low-resolution hotspots into contour line representations with smooth boundaries, increasing generation speed \cite{johansson2015crime}. Other tools such as non-parametric density estimators for creating maps that are smooth and accurate \cite{chainey2013gis,eck_mapping_2005}, hotspot optimization tool (HOT), made data mining techniques for hotspot mapping easier to implement \cite{Wang:2013:CHM:2584128.2584145}. Prior research has shown that these tools and algorithms may be used to extract information on criminal organizations from large real-world crime datasets, particularly police-reported crime data, which is very difficult to achieve using standard crime analysis approaches \cite{6425708,chainey2013gis,1193063}. 

However, recent SIGCHI research pointed out how such tools and approaches have been linked to several concerns \cite{brayne2017big}. According to research, racial bias can be found in police reports including racial profiling of vehicles \cite{a, doi:10.1080/10439463.2018.1473978, doi:10.1080/21565503.2016.1160413, doi:10.1111/j.1745-9125.2006.00061.x}, pedestrian stops \cite{fagan2016stops,goel_precinct_2016,noauthor_down_nodate}, traffic fines \cite{dunn2009measuring}, narcotics enforcement and arrests \cite{lynch2013policing,kochel2011effect,beckett2006race}, use of force \cite{buehler2017racial,legewie2016racial,nix2017bird} etc., which can be further amplified if this data has been fed to the prediction tools. The concern is that racially biased police practices will be oriented toward particular places rather than others and knowing that they are in a prediction area may increase police officers' awareness in ways that accentuate biases such as a minority being subjected to discriminatory police actions  \cite{ferguson2012predictive}. Along with racial bias, privacy concerns have also been raised. \cite{article457,34}. Because police reports are public information and several police departments offer web-based crime mapping tools on their websites, disseminating spatial crime data might be difficult when the locations of crimes can be linked to specific residences and, thus, specific individuals \cite{34}. These issues have influenced recent research trends aimed at identifying possible prejudice and unethical implementations.

\subsection{Algorithmic Decision-Support(ADS) Systems in the Public Sector}

Algorithmic systems are increasingly being deployed across several public sector organizations such as child welfare \cite{cheng2022child, kawakami2022improving, saxena2020child}, public education \cite{mcconvey2023human, robertson2021modeling}, homeless services \cite{moon2024human, kuo2023understanding, kube2022just}, criminal justice \cite{stevenson2022algorithmic}, unemployment services \cite{holten2020shifting}, and welfare benefits \cite{eubanks2018automating} to make high-stakes decisions about citizens’ lives. For instance, algorithms are being used in the criminal justice system to determine sentencing length \cite{grgic2019human}, allocate resources to neighborhoods \cite{chainey2013examining}, and predict recidivism \cite{pruss2023ghosting}. In child welfare, algorithms predict the risk of future maltreatment \cite{cheng2022child}, determine suitable caregivers \cite{saxena2022unpacking}, and decide on family services \cite{saxena2023rethinking}. Public education relies on algorithms for student zoning and performance evaluation \cite{robertson2021modeling}. Job placement centers use algorithms to profile job seekers and make placement decisions \cite{holten2020shifting, ammitzboll2021street}. Additionally, algorithms establish eligibility criteria and distribute benefits to families in need \cite{eubanks2018automating}. The deployment and use of algorithms in these sectors have far-reaching consequences, often impacting lives and, in some cases, resulting in life-altering or life-and-death outcomes. As a result, SIGCHI researchers have become increasingly interested in understanding how human-AI decision-making unfolds in these domains \cite{kawakami2022improving, saxena2021framework}, the perspective of frontline workers whose labor is impacted by these technologies \cite{cheng2022child, ammitzboll2021street}, the impact of organizational and policy-related factors \cite{saxena2023algorithmic}, as well as unpacking implications for the worker-centered design of ADS in these contexts \cite{saxena2022unpacking, kawakami2022care}.

For instance, researchers have studied how caseworkers in child welfare collaboratively engage with AI systems and how AI-assisted decisions are made within the bounds of organizational and legal constraints \cite{kawakami2022improving, saxena2021framework, holten2020shifting}. In a similar vein of work, Kawakami et al. \cite{kawakami2023training} develop the notion of \textit{critical use} that centers the practitioners’ ability to situate AI predictions against the knowledge that is uniquely available to them, i.e. - pre-emptively recognizing that an information asymmetry between the AI and the human is likely to exist and training workers regarding the appropriate reliance on AI systems. Researchers have also explored validity considerations from the social sciences to help improve machine learning problem formulation and assess whether the use of an AI system in a given context is appropriate to begin with \cite{coston2023validity, kawakami2023recentering}. Similarly, Guerdan et al. \cite{guerdan2023ground} reviewed several case studies of AI systems in the public sector and developed a causal framework to identify sources of target variable bias and their impact on human-AI decision-making. In addition, researchers have also explored opportunities to support collaboration in human-AI decision-making \cite{buccinca2021trust, lai2020chicago, green2019disparate} by developing explanations \cite{bussone2015role}, training protocols \cite{kawakami2023training}, and prescribing away from predicting outcomes \cite{saxena2023algorithmic}.

In sociotechnical systems, however, AI interfaces rarely contain a clear feedback mechanism that allows users to comprehend the effects of their actions on the system. The growing prevalence of opaque algorithms deployed within public agencies raises concerns about users' and the community's education regarding their operation, adoption, and potential biases embedded in them \cite{eslami_understanding_2017}. We add to the body of knowledge by examining the interaction between a crime-mapping algorithm and users from various backgrounds to identify opportunities and challenges in designing human-centered algorithms that must operate in the public's interest.

\subsection{Community Members' Participation in the AI Deliberation Process}

Academics and journalists have gathered substantial evidence highlighting the outcome-oriented harms resulting from algorithmic systems and have specifically drawn attention to disparate harms inflicted on impacted communities \cite{raji2022fallacy, shelby2022identifying,johnson_2023, Villasenor_2021, garance_2022, garza_2022}. Consequently, the design process of AI systems is often criticized for not involving community members in the process even though they are the ones who face the consequences of algorithmic decisions. As a result, researchers have argued for expanding participation in the AI deliberation and design process to incorporate a variety of stakeholders and elicit their feedback at different stages of the process \cite{delgado2023participatory}. SIGCHI researchers have conducted studies that elicit community members’ perceptions of fairness \cite{van2021effect, van2019crowdsourcing, srivastava2019mathematical} as well as designed frameworks to elicit stakeholders’ nuanced notions of fairness regarding algorithmic systems deployed in real-world contexts \cite{cheng2021soliciting}. Similarly, researchers have engaged in co-design processes with workers to develop AI interventions that are centered in their practice and foster worker well-being \cite{holstein2019co, zhang2022algorithmic, spektor2023designing}.

Regarding algorithmic systems deployed in the public sector, researchers have closely worked with civil servants whose labor is impacted by these systems and elicited their feedback to inform the ethical and human-centered design of AI systems \cite{brown2019toward, cheng2022child, saxena2020conducting, ammitzboll2021street}. Recently, SIGCHI researchers have contributed to this literature by working directly with impacted community members to understand their perspectives and needs regarding the use of AI systems in public agencies and NGOs. For instance, Lee et al. \cite{lee2019webuildai} developed a participatory framework that facilitates ML model development with a variety of stakeholders and improves distributive outcomes and procedural fairness. Brown et al. \cite{brown2019toward} conducted participatory workshops with community members to understand their perspectives on the use of a predictive risk model (PRMs) deployed within the child welfare system. Stapleton et al. \cite{stapleton2022imagining} build upon this work by conducting design workshops with parents investigated by the child welfare system and social workers to question whether such models should be developed, to begin with, and found that both these stakeholders had significant concerns regarding PRMs and instead wanted to address the problems motivating the use of such systems. Similarly, Kuo and Shen et al. \cite{kuo2023understanding} engaged stakeholders, including service providers, county workers, and unhoused individuals, in \textit{AI lifecycle comicboarding} sessions for homeless services. Their findings indicate that stakeholders can actively contribute to the early stages of the AI deliberation process, influencing problem formulation, dataset selection, and modeling choices. This study further builds upon this body of research by introducing an intuitive AI application that acts as a boundary object that the participants can interact with, helps facilitate context-specific and nuanced conversations with them about policing practices in their city, the use of AI systems by police departments, and elicit their feedback regarding the ethical design and use of such systems.


\section{Methods}

We conducted a mixed methods study with participants from Milwaukee, Wisconsin, a mid-sized, midwestern city in the United States from May to August 2019. A total of 60 community stakeholders participated in this study and helped us examine how people from different educational and professional backgrounds interact with crime mapping algorithms. Based on prior literature about common crime mapping algorithms used in police departments \cite{chainey_utility_2008}, we developed an interactive application that displays a heatmap output of a standard Kernel Density Estimation (KDE) algorithm given crime data and particular user inputs. Our study consisted of 2 parts - \textbf{1)} through an in-lab activity, we examined participants' interactions with the crime mapping algorithm where they chose different parameters for the KDE algorithm, interpreted the results, and identified hotspots on the map. After the lab activity, participants completed the NASA-TLX survey which helped us assess their self-reported mental workload as a result of the activity, \textbf{ 2)} we conducted semi-structured interviews with the participants to gather their perspectives and needs regarding the ethical design and use of such systems by police departments. This study was approved by the Institutional Review Board (IRB) of our university in the United States.

\begin{figure*}[t!]
\centering
\includegraphics[width=0.7\textwidth]{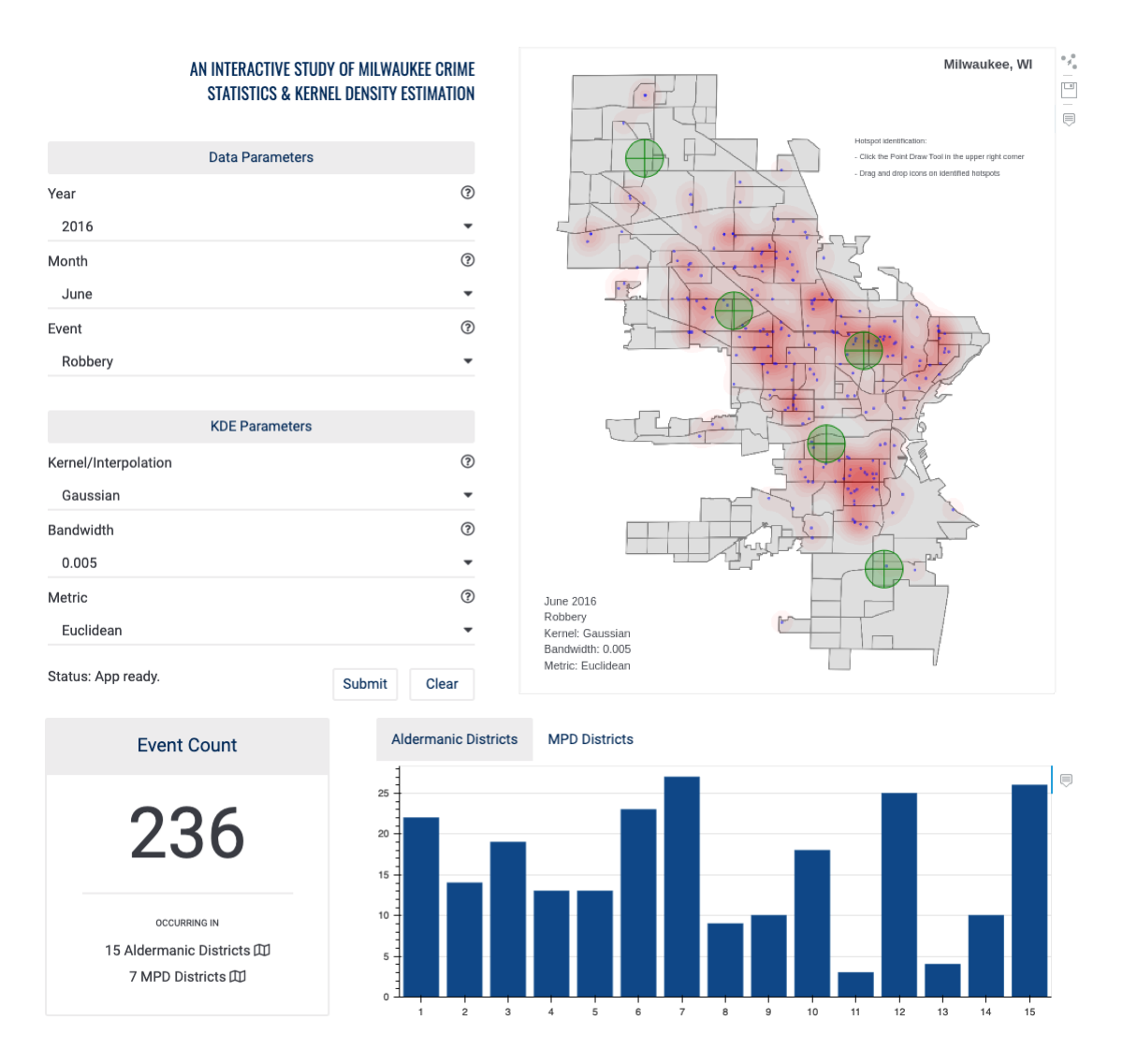}
\caption{\textbf{Screenshot of the Algorithmic Crime Mapping Application}}
\label{fig:appResponse}
\end{figure*}

\subsection{Application Design}
Crime data used in our application was taken from the Wisconsin Incident-Based Reporting System (WIBRS), a publicly accessible database organized by the City of Milwaukee and Milwaukee Police Department \cite{wibrs}. Data had to be accessed by time and was collected starting with 2017 (the last full year before this project began) and then worked backward. We were able to curate a dataset of crimes from 2012 to 2017. We chose to focus on 3 distinct quality-of-life crimes - robbery, larceny, and motor theft because they are most commonly reported and tend to have the most effect on an average person's life \cite{james2008crime}. Moreover, city residents would also be familiar with or might have encountered them. In contrast, rape, murder, terrorism, etc. receive significant media coverage but are comparatively rare events and also disproportionately reported between rural and urban settings \cite{livingston1997police}. Street addresses provided in the dataset were then cleaned using a manually assembled regex library. Then they were geocoded using Google's geocoding service, and organized into individual months.

To facilitate our study, we built an online, interactive application (Figure \ref{fig:appResponse}) that displays the heatmap output of a Bivariate Kernel Density Estimation (KDE) based on specified data and KDE parameters. We chose to present KDE outputs for two reasons: first, we knew from prior knowledge of common practices by crime labs that KDEs are one of, if not the most commonly used crime mapping algorithm \cite{haque2019exploring}, and second, we wanted to examine participants' interactions with and interpretation of the KDE heatmap. Each parameter dropdown list had a question mark tooltip to answer some general questions about parameters. The data parameters allow users to choose one type of crime from a specific month from 2012 to 2017. To the extent possible, we chose to develop the design features of our application that closely resemble popular crime mapping platforms used in police departments. Moreover, such crime mapping platforms are usually not accessible to researchers so we decided to create an in-house application that closely mimics a real-world algorithmic crime mapping application. KDE parameters and options are outlined in Table 1. 

\begin{table}
    \small
    \begin{tabular}{|>{\raggedright}p{1.8cm}|>{\raggedright}p{1.9cm}|>{\raggedright\arraybackslash}p{3cm}|}
    \hline
    \textbf{KDE parameters} & \textbf{Default Options} & \textbf{Other Available Options} \\ \hline
    Kernel & Gaussian & Tophat, Epanechnikov, Linear \\ \hline
    Bandwidth & 0.005 & 0.0025, 0.0075, 0.01 \\ \hline
    Distance Metric & Euclidean & Manhattan, Chebyshev \\ \hline
    \end{tabular}
    \caption{List of Interactive KDE Parameters and Options}
\end{table}


\subsection{How does KDE Map Crime?}

KDEs attempt \cite{backurs2019space} to find and represent the underlying probability distribution function (PDF) that a set of data was taken from. By providing a PDF, the heatmap allows users to predict high- and low-probability areas for future events. KDEs accomplish this by smoothing each discrete data point into a two-dimensional probability distribution function with the original point at the mean and aggregating the PDFs into a singular heatmap for the entire area. The shape of the distribution that data points are smoothed into is defined by the \textbf{kernel parameter,} also called the interpolation method. The \textbf{bandwidth} parameter controls the width of each distribution. In statistics, this would be analogous to the variance of a symmetric distribution. For example, a Gaussian kernel with a bandwidth of one will result in the standard normal distribution, though generally a much smaller bandwidth is preferred in order to produce meaningful results. Higher values for bandwidth result in much smoother outputs, which would lose power but reduce bias from overfitting. The \textbf{distance metric} parameter controls how the distance between points is measured. The most commonly used distance metric is Euclidean, or merely a straight line between two points on a plane, which is the default setting for the distance metric. In our application, after submitting an initial set of parameters, users are shown the calculated heatmap with several interactive features, as well as a bar chart displaying the number of events in either each police district or each aldermanic district. Hovering over data points in the heatmap shows a pop-up with information about the date, district, and location of the event.

\vspace{-0.2cm}
\subsection{Participants and Data Collection}
We decided to recruit participants local to this midwestern city as most empirical studies on algorithms in criminal justice either deal with historical data about people suspected or convicted of crime or crowd workers (e.g. Amazon Mechanical Turk, Crowdflower, etc.) \cite{barabas2019technical, rubya2021comparing, sen2015turkers, lee2021included}. Local residents are able to situate the results and interpret them in the context of their lived experiences allowing us to better understand their perspectives and needs. 

Recruitment was undertaken in the form of advertisement flyers posted around the city in coffee shops, community centers, and public libraries as well as various online forums and groups. As much as possible, we tried to disperse our recruitment efforts through all parts of the city including high- and low-income areas as well as those parts of the city with historically higher or lower rates of crime. To recruit LEA participants, we made special efforts to reach out to various police departments in the county which included about 8 different police departments. While we were successful in recruiting LEA participants, this was significantly hard because LEAs in these departments are busy, under-resourced, and generally reticent to speak to researchers as there is a history of critical research and policy findings about the practices of police departments \cite{july15july20}. In the end, we ended up with participants (\(n= 60\)) with varying academic and professional backgrounds. An overview of the participants' demographics can be found in Table 2. We were able to roughly align gender and education with the 2010 US Census \cite{bureau} but were unable to do so for age. Thus, our sample age distribution skews about 20\% younger than in the 2010 US Census but is a more relevant and local sample than prior work.  Each participant was given a specific participant identification number to allow us to use their data and interview transcript for analysis without revealing personally identifiable information.

\subsection{Participant Grouping}
We assigned participants to one of the three groups depending on their self-reported background and experience. \textit{Group 1 (\(n_1\) = 39)} consists of those with no background relevant to algorithmic crime analysis. These are mostly area residents with white or blue-collar jobs in local, food services, finance, and healthcare industries. We refer to them as \textbf{community members}. \textit{Group 2 (\(n_2 = 14\))} participants have a technical background involving software engineering or data science. We refer to them as \textbf{technical participants}. \textit{Group 3 (\(n_3 = 7\))} consists of law enforcement agents (LEAs) who work professionally for a police department. These participants also have prior experience working with algorithmic crime mapping and we refer to them as \textbf{LEAs}.

\begin{table}[]
\small
\begin{tabular}{|>{\raggedright}p{1.7cm}|>{\raggedright\arraybackslash}p{6cm}|}
\hline
Demographic Criteria & Participant Description (with number of participants)                                                                                                \\ \hline
Gender               & Male \textbf{(23)}; Female \textbf{(36)}; Trans male \textbf{(1)}                                                                                                            \\ \hline
Age range & 18-21 years \textbf{(29)}; 22-30 years \textbf{(20)}; 31-40 years \textbf{(4)}; 40+ years \textbf{(7)}                                                                                                        \\ \hline
Education            & High  school diploma or GED \textbf{(2)}; Undergraduate \textbf{(32)}; Bachelors \textbf{(20)}; Masters \textbf{(5)}; Doctorate \textbf{(1)}        \\ \hline

Job type             & Full-time \textbf{(17)}; part-time \textbf{(14)}; unemployed \textbf{(4)}; self-employed \textbf{(3)}; full-time student \textbf{(21)}; retired \textbf{(1)} \\ \hline
\end{tabular}
\caption{An overview of the participants' (n=60) demographics}
\vspace{-0.5cm}
\end{table}


\subsection{In-Lab Activities}
To gauge the participants' ability to understand, interpret, and use algorithmic crime analysis, we asked them to perform a series of activities on heatmaps of three different complexities (Maps A, B, C). Each session was approximately 60 to 90 minutes long. Initially, participants were presented with a consent form (approved by the authors' institutional IRB) to sign, indicating their agreement to share their data for research purposes. The form also emphasized their right to withdraw from the study at any time and choose not to share their data. Next, they were provided with a pre-interview questionnaire containing demographic-related questions, including information about race, gender, education, employment, socio-economic status, and so forth. After that, we provided a brief overview to the participants about the activities as well as explained the different parameters to them. While participants were able to change the parameters of the KDE to one of several options, data parameters were provided and consistent across maps. Therefore, the complexity of the underlying data patterns rather than a specific heatmap output was considered when choosing exemplar data parameters for each complexity. To determine the complexities of which maps would be shown, the research team went through a rigorous process of filtering through and choosing different heatmaps collaboratively and selected those with clearly different levels of complexity ranging from distinct hotspots scattered across the map to hotspots that blurred together and were not easy to pinpoint.
Map A is the least complex, exhibiting distinct clusters. Map B is more complex, with blurred edges around clusters. Map C is the most complex, with clusters blending but still showing slight density variations. All co-authors identified and agreed on the 4 data points for each map. Next, participants were asked to estimate minimum and maximum hotspots in each presented heatmap.

Following this, participants were asked to envision green circles with crosshairs as the optimal patrol areas for one police unit. They were then tasked with specifying the minimum and maximum number of circles required to address crime effectively based on the heatmap. Finally, they placed 5 circles in locations representing the most efficient allocation of resources. An example heatmap can be seen in Figure \ref{fig:appResponse}. The session, led by one author and assisted by another, involved these interactive lab activities. Post-session, the authors collaborated to refine observation notes and shared these impressions. Participants then completed the NASA-TLX survey \cite{hart1988development} to evaluate mental workload, a standard instrument for assessing the efficacy of a system in HCI \cite{NASA}. Participants then answered a survey with questions about their prior level of familiarity with algorithmic crime analysis, their understanding of the application, feelings towards law enforcement and government, and general demographics. Finally, participants were interviewed with a set of questions in a protocol based on their familiarity with such algorithmic systems, their views on the legality, ethics, and fairness of data, and their concerns regarding algorithmic crime mapping. The interview sessions were audio-recorded with participants' consent.

\vspace{-0.2cm}
\subsection{Analysis}

We calculated the probability distribution of using the default parameters to determine the default behavior among the three groups. Additionally, we used the NASA-TLX survey to assess the mental workload of interacting with the crime map. The TLX involves participants reflecting on the task and making 15 paired comparisons across six dimensions to gauge workload. For instance, they needed to decide whether \textit{Performance} or \textit{Frustration} represented the more important contributor to the workload for the specific task they had recently performed. The second step involved participants rating each of the six dimensions on a Likert-type scale. The raw score for each of the six items is multiplied by the weight from step 1 to generate the overall workload score per task.

Qualitative interviews were transcribed using the online platform \textbf{Otter} \cite{otter}. Subsequently, two authors manually reviewed the automated transcriptions to ensure accuracy. Thematic analysis \cite{braun2006using} with constant comparison was employed, and two authors conducted iterative coding of textual codes. A third co-author verified the coding process. These low-level codes were combined into categories, and finally, these categories were collated into higher-level themes. These themes are presented throughout our results below with appropriate quotes and explanations. 


\section{Results}

\begin{figure*}[h]       
    \fbox{\includegraphics[width=0.31\textwidth]{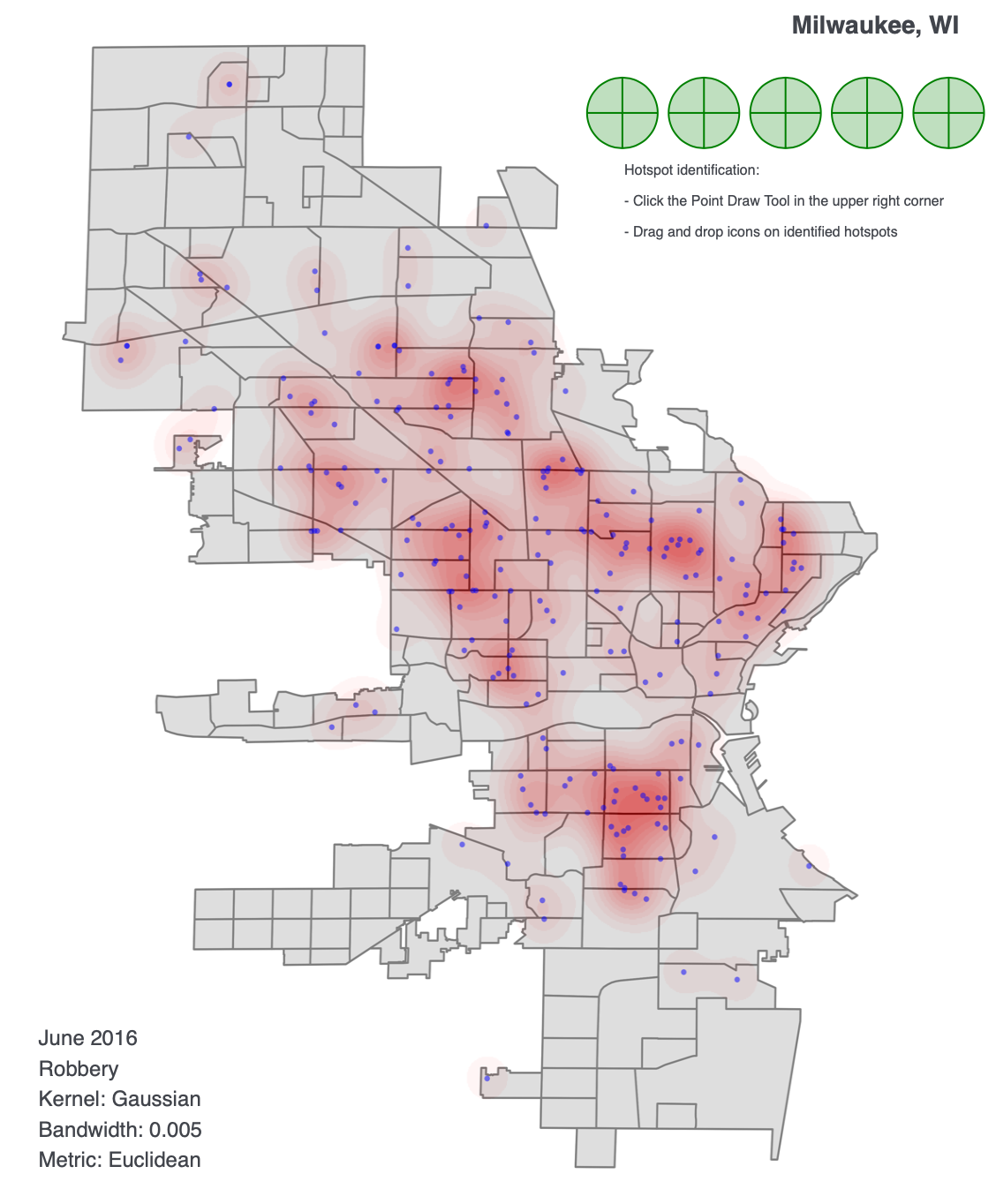}}   
    \hspace{0px}
    \fbox{\includegraphics[width=0.31\textwidth]{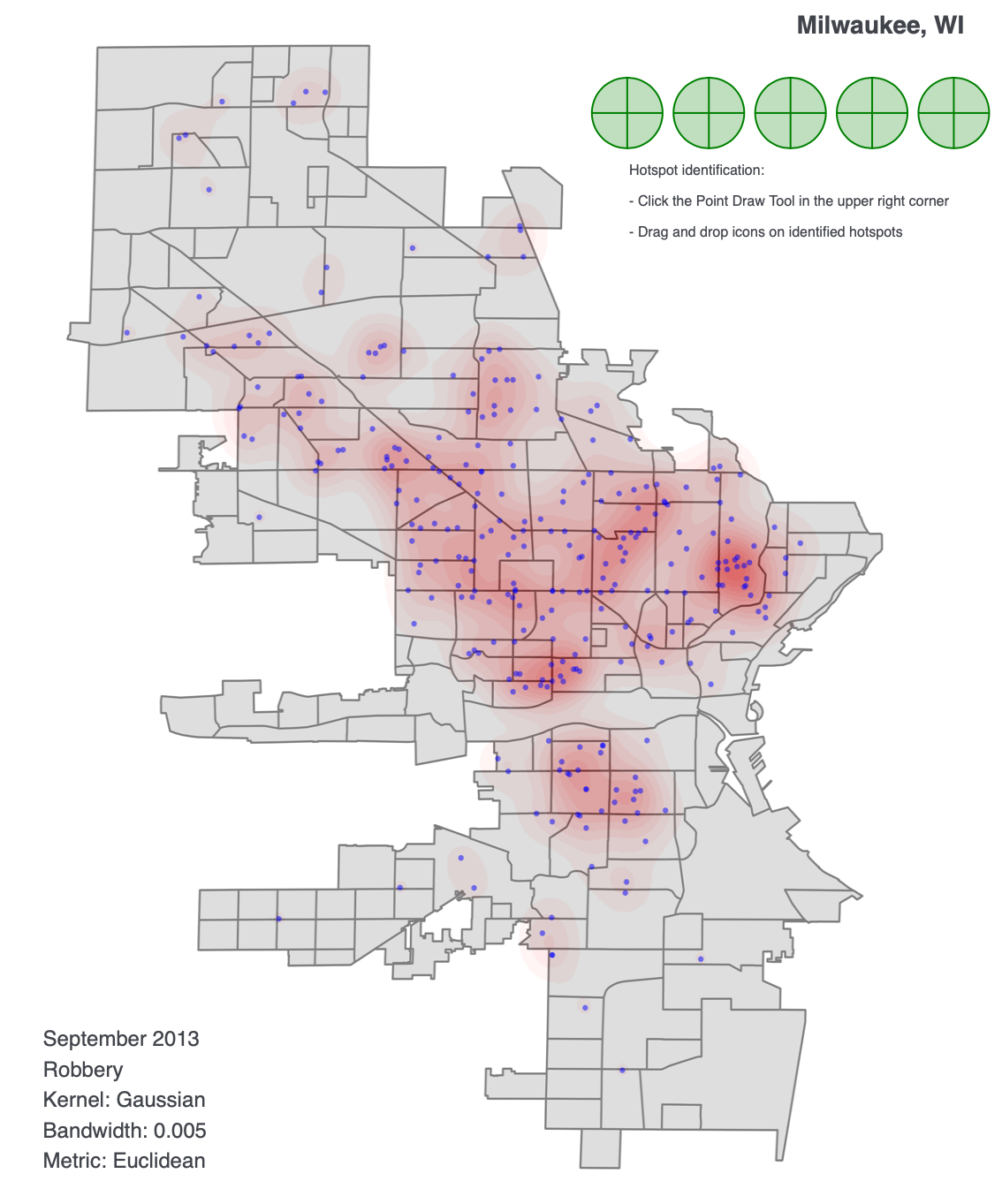}}
    \hspace{0px}
    \fbox{\includegraphics[width=0.31\textwidth]{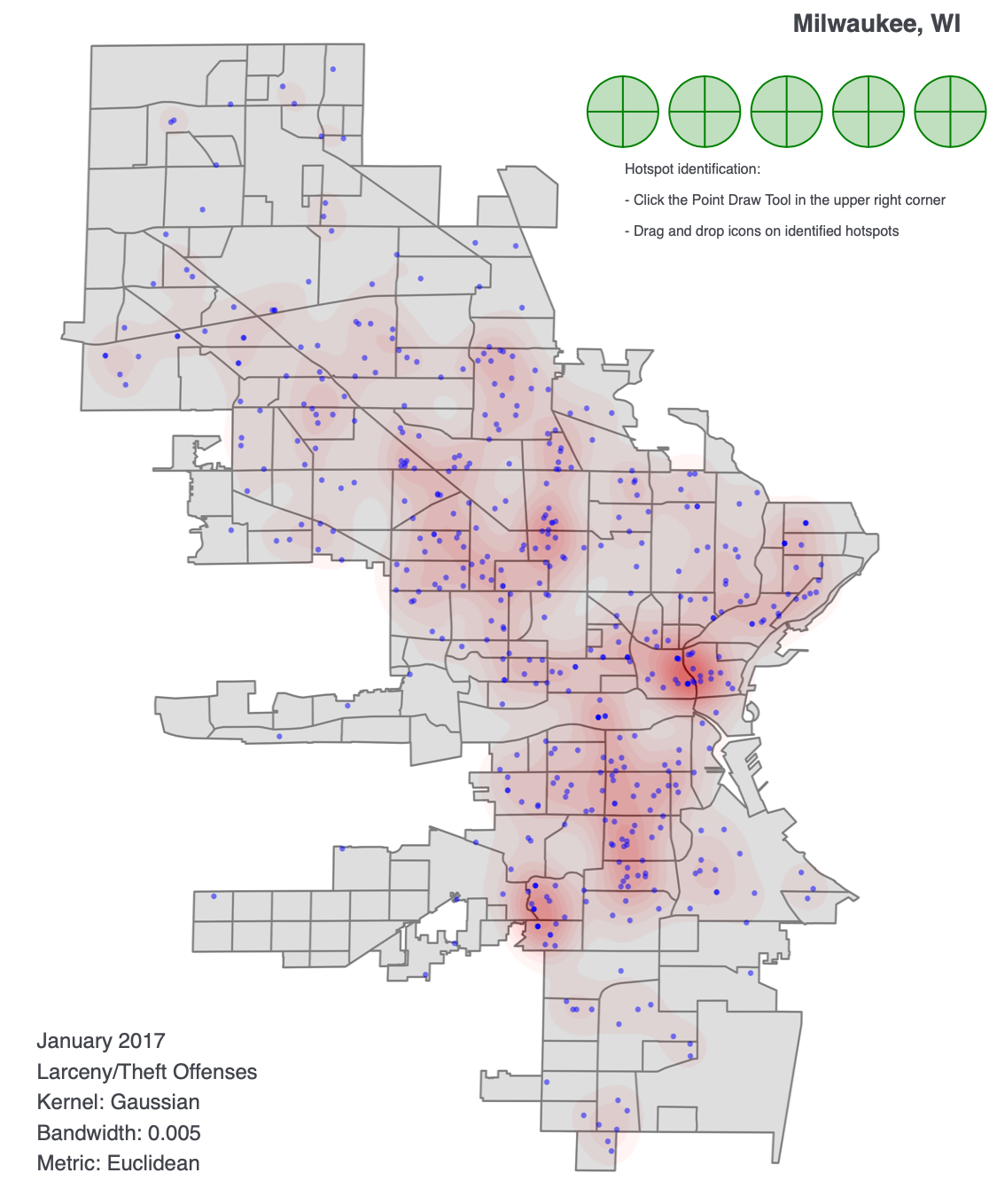}}
    \caption{Algorithmic Crime Maps With Increasing Level of Complexity from Left to Right.}
    \label{materialflowChart}
    \vspace{0.2cm}
\end{figure*}

Before engaging in the lab activities, we provided the participants with an overview of algorithmic crime mapping. We shared with them how the application was expected to work and the general intent behind algorithmic crime mapping (i.e., efficient allocation of resources on the part of the police department). We also showed them how they could produce different maps by changing the parameters. After observing their interactions with the application, we conducted semi-structured interviews to gather their impressions of this technology and learn about their needs. To improve the readability of the results, we first share participants’ perspectives (RQ1), followed by the findings from participants' interactions with the applications (RQ2), and finally, their needs regarding a community-centered vision of this application (RQ3).

\subsection{Participants' Perspectives on Algorithmic Crime-Mapping (RQ1)}

\subsubsection{Community members found utility in algorithmic crime mapping but were concerned about the motivations driving the use of such technologies.}
Most community members and technical participants shared their skepticism about crime mapping where they were primarily concerned that such technologies could lead to over-policing of specific regions, used as a revenue generator by targeting specific areas, and create a feedback loop that leads to over-policing of specific neighborhoods and communities. Several participants questioned the motivation behind crime mapping and wondered if the technology was being used to protect citizens and deter crime or whether it was being used simply to identify a high-crime region and target it to generate revenue. For instance, participant C1 asked: 

\begin{quote}
    \small{"I think this is like the idea where you're putting the police where they can make the most money versus actually control crime and danger. Maybe there's a stretch of highway where it [speed] goes from 70 to 55. And so you are going 75. And they're trying to catch people who are not speeding with speed traps. So you're going to catch a lot of people there. But is that the most efficient way to, you know, have order and justice in a city? Probably not." - C1, Community Member}
\end{quote}

Here, another participant also brought up the example of speed traps as well as other traffic violations and believed that traffic cameras and other forms of automation were better suited to deter those crimes than increasing the police presence in the neighborhood as well as questioned whether increased police presence led to safer neighborhoods. The participant who lives close to an urban college campus shared: 

\begin{quote}
    \small{"I grew up in a little town, where we actually knew all the cops, and if something did happen, you probably knew exactly who it was. It was a little different here in the city. They say,  we have a big police department, security cameras everywhere, and then private campus police. And it still kind of feels like a lot of times it's still not enough. And so I guess it's really important that it's not just the perception of safety, but we're actually having like the data on how safe we are, how much we're actually doing to stop crime and how much crime is actually happening." - C14, Community Member}
\end{quote}

This concern was also echoed by community member C2, who shared that they believed that such technologies could be used to target and set up people in low-income urban areas.

\begin{quote}
    \small{"I mean, it's like you're setting people up that don't really deserve to be set up. Like you're looking for the crime, but it's not really a crime for me [referring to stealing], like it should be all about violent crimes or harassment or things that will actually like hurt an individual." - C2, Community Member}
\end{quote}

Some technical participants were also concerned about how the application changes over time and whether focusing on certain crimes draws attention (and resources) away from more serious crimes as well as the long-term impact of increased police presence in some neighborhoods. For instance, T5 shared:

\begin{quote}
    \small{"It's kind of a feedback loop.... You took it [crime data] on more people at one location and you put more cops there. You're going to get more tickets there. And now you're going to put even more cops there. And then, it's the same thing over and over. And are you really helping?" --T5, Technical Participant}
\end{quote}

On the other hand, several community members and law enforcement officers found utility in the tool. Community members appreciated that such applications could provide a real-time snapshot of crime prevalence in different neighborhoods and help them make important decisions such as purchasing a house and finding the right school district for their children. Participant, C6, went so far as to say that they would be willing to freely share their data to create such a public resource.

\begin{quote}
    \small{"I would share my data, pretty open book, and you know, they need up-to-date current information for everybody. Otherwise, if you're just using old statistics and opinions, your methods and results aren't going to keep up with all the constant change that's happening. Especially as our generation is now the one that will start being buying homes and having children and like, we're going to be the ones that are going to start living in these neighborhoods. And so it's important to know, it is a concern for the data being up to date." -- C6, Community Member}
\end{quote}

Law enforcement officers also shared that they found utility in such a tool because it helped them make real-time data-driven decisions based on crime statistics in certain areas and decide how many patrol cars they needed on the streets. For instance, L8 shared: 

\begin{quote}
    \small{"Oftentimes, we take a look at our calls for service, How many calls for service do we have for a specific time period? The entire day? And that determines how many officers we would have out at first, as opposed to the second shift. So for example, we take a look at 12am until 3am, and we have a high population of disorderly conduct at this location, then it makes sense to send officers to that location to hopefully make sure that the incidents aren't happening. Because usually, the visibility of police will lower activity." --L8, Law Enforcement Agent}
\end{quote}

In sum, we learned that several community members found utility in algorithmic crime mapping but were primarily concerned about the motivations driving the use of such tools. They shared that instead of protecting people and deterring crime, the tool might become a way for police officers to “find crime” and set up people as a means for writing more tickets. Here, it is important to note that the distrust in policing practices underscored the majority of our conversations with community members who often referred to prominent cases of police brutality to highlight police wrongdoing. As we highlight in Section 4.3, several community members brought up examples where the police reports did not match the video evidence shared on social media by bystanders.

\subsection{Participants' Interactions With Algorithmic Crime Mapping (RQ2)}
We conducted in-lab activities with our participants where we first gave them an overview of the application and explained to them how they could change parameters to produce different crime maps. Next, we asked the participants to estimate the number of hotspots on the maps based on their knowledge of the city. 

\subsubsection{Law Enforcement Agents (LEA) significantly over-estimated hotspots.}
We conducted statistical tests to assess the differences in the estimation of hotspots across the three different groups (i.e., community members, technical participants, and LEAs) and the three different maps. Please see \hyperref[firstappendix]{Appendix A} for more details about the statistical tests. The intent behind this activity was to assess participants’ perceptions about the prevalence of crime based on their background and whether it impacted how they interacted with the application.

We learned that LEAs used their hyperlocal knowledge of the different intersections (within neighborhoods) in the city to identify epicenters for criminal activity, however, in doing so, significantly overestimated the number of hotspots. For instance, an LEA who was interacting with the crime map shared that he was looking at all the major intersections in the vicinity of hotspots and tying criminal activity to his street-level knowledge of those intersections:

\begin{quote}
    \small{"With property crimes, such as burglaries. I have to take a look at the way the map is as well. I know, for example, [anon] Avenue on the North side is a main thoroughfare.... A lot of our robbery offenders do tend to reside on the north side." -L7, Law Enforcement Agent}
\end{quote}

\vspace{-0.1cm}
On the other hand, community members' and technical participants’ estimations were much closer to the optimal number of hotspots (as predicted by KDE). Here participants used their local knowledge of different neighborhoods to conceptually place the hotspots and verbally explained which neighborhoods (not specific intersections) they would place the hotspots in. We also witnessed them validating their interpretation with their prior knowledge of specific neighborhoods. P14 mentioned how he had experience with a similar application before, however, his decision to place hotspots was also influenced by his familiarity with specific neighborhoods.

\begin{quote}
    \small{"I have a similar experience with software that shows the heat map [of crime data]... so I am capable of making the decisions based on the parameters I think ... but you know, I know my neighborhood. I know where to put it [the hotspots] or not .. "- P14, Technical Participant}
\end{quote}

\vspace{-0.1cm}
LEAs could have been overestimating hotspots for several reasons. Evidence in criminology literature indicates that police officers may overestimate the prevalence of crime in certain neighborhoods or communities \cite{buil2021accuracy}. Here, the perception of crime can be influenced by several factors such as the nature of policing work, media coverage of crime, and perceptions about low-income and minority communities \cite{smith2007explaining}. There is also evidence that indicates that micro-level crime analysis as undertaken by LEAs (i.e., assessing specific intersections) is prone to a larger risk of selection and reporting bias as compared to crimes aggregated and assessed at a larger scale. However, in our conversations with the LEAs, we also learned that they did not draw clear distinctions between violent crimes and quality-of-life crimes. They often referred to the fact that different types of crimes can be interlinked and at times quality-of-life crimes can escalate to more serious crimes. That is, they drew on their domain expertise to focus on several hyperlocal hotspots instead of a general hotspot covering a neighborhood that a squad car could easily patrol. This finding was further supported by the next activity where we asked participants to place green circles on the map that demarcate the area that one patrol car should cover.

\vspace{-0.1cm}
\subsubsection{Reasons for Placing Patrol Cars - Deterring Crimes versus Ensuring Quick Availability of Backup.}

For the next activity, we provided participants with five green circles (or crosshairs) with each circle depicting the area a patrol car could easily cover. Here, the participants could see the hotspots depicted on the map and had to decide where and how to allocate these resources. Most community members and technical participants placed the green circles further apart and tried to maximize the area being covered such that most neighborhoods would receive some police presence. Participants justified this choice by stating that they intended to deter crime by ensuring police presence. We also noticed several participants doing some mental calculations where they started to guess how long it would take a police car to patrol the green circle and placed other green circles based on that.

On the other hand, we witnessed that LEAs placed the green circles closer together and also asked for more green circles because five circles were insufficient for covering the hotspots. While the community members and technical participants prioritized spreading the green circles across the map to cover a wider area, LEAs placed them closer together to ensure more police presence in a specialized patrol zone and the quick availability of backup for a patrol car if a situation were to arise. They also used their experience in crime analysis and described the map as to why they placed the green circles in a certain fashion. L4 first explained to us their definition of a hotspot as a current area where a patrol officer already is and used that in interpreting the map.

\begin{quote}
    \small{"So the people, the officers who are the administrators who would be doing data analysis or looking at crime maps, might give directive of having a specialized patrol zone, and to make sure that officers were in that patrol zone." - L4, Law Enforcement Agent}
\end{quote}

\vspace{-0.1cm}
Several LEAs touched upon the theme of deterring crime by increasing police presence, however, as depicted by the exemplar quote above, their decision-making is impacted by several real-time factors (e.g., the creation of a specialized patrol zone due to increased criminal activity as well as ensuring that patrol cars receive timely support).

\vspace{-0.1cm}
\subsubsection{Participants Experimenting With Different Maps - Community members and technical participants were more inclined to generate new maps than LEAs.}
During our observations of participants’ interactions with different crime maps, we recorded whether participants changed the parameters to create different maps and their responses to these different maps. Please see \hyperref[secondappendix]{Appendix B} for results of the quantitative analysis. Both technical participants and community members interacted with parameters and generated different maps. Some participants gamified the process for themselves and started to guess how changing certain parameters would affect the map. Technical participants shared that they had seen static geographical heatmaps before and knew how to generate them by writing code but had never interacted with an interactive map. Participant T9 shared:

\begin{quote}
    \small{"When I did it the first time, it was the default parameters, and I wanted to see, basically, what sort of difference it would make using the other sets. I’ve heard of a few of these parameters… So I figured the best way would just be to run it and see what actually happens." -- T9, Technical Participant}
\end{quote}

\vspace{-0.1cm}
Several community members appreciated the opportunity to interact with such an application and understand how resources might be allocated using such tools. They shared that such transparency, even though carried out as a lab activity, helped them develop trust. For instance, a community member shared: 

\begin{quote}
    \small{"This is really important, I think. I mean, I, personally like to know what's going on [in regard to crime], especially if I'm living in the area and paying taxes on property, and in general, to have an idea of what's going on and why that's going on? I think this would be good for any person." -- C15, Community Member}
\end{quote}

\vspace{-0.1cm}
Several community members also shared that they liked the easier maps because those maps provided them with bigger circles (i.e., hotspots) and gave them a quick overview of the neighborhoods. They shared that these would be a useful communication and awareness tool for the community.

The results, regarding LEAs, were quite surprising for the research team. We had anticipated that LEAs would be most likely to interact with the parameters, generate different maps, and connect the findings to street-level occurrences over different periods. However, none of the LEAs changed parameters and spent a significant amount of time interpreting the first map presented to them. Instead of using the crime maps to augment their knowledge of geospatial and temporal characteristics of different crimes, LEAs quickly started connecting their street-level domain knowledge to the hotspots on the map, i.e. - the first piece of information presented to them. Instead of using the tool as a decision aid that would augment their knowledge, they used it as a means to validate what they already knew about certain neighborhoods, sections of highways, and intersections. When urged to change parameters and create a different map, an LEA who had experience working with crime maps stated: 

\begin{quote}
    \small{"This is okay [the default map]... the algorithms are usually designed in such a way that it will give you some satisfactory outcomes. Usually what you need to think about is what prompts them to make that decision. Data that goes into it is very important" -- L2, Law Enforcement Agent}
\end{quote}

\vspace{-0.1cm}
The LEA makes an interesting point here regarding the data that we further delve into in the next subsection. However, for the most part, we witnessed that LEAs exhibited a strong anchoring bias where the default map (i.e., the `anchor') provided a `satisfactory outcome' that seemed to align with what they already knew about the city. Anchoring bias is a cognitive bias that refers to the tendency of practitioners to rely too heavily on the first piece of information encountered when making decisions and has been observed across several domains \cite{furnham2011literature, azzopardi2021cognitive, saposnik2016cognitive}. However, as we discuss in the next section (see Section 4.3), the LEAs asked for features (e.g., the ability to draw districts on the map, see spatiotemporal relationships between different types of crime, and use information from on-the-ground investigations) that would have allowed them to derive more context-specific information.


\begin{figure}
    \centering
    \begin{subfigure}{0.35\textwidth}
        \includegraphics[width=\textwidth]{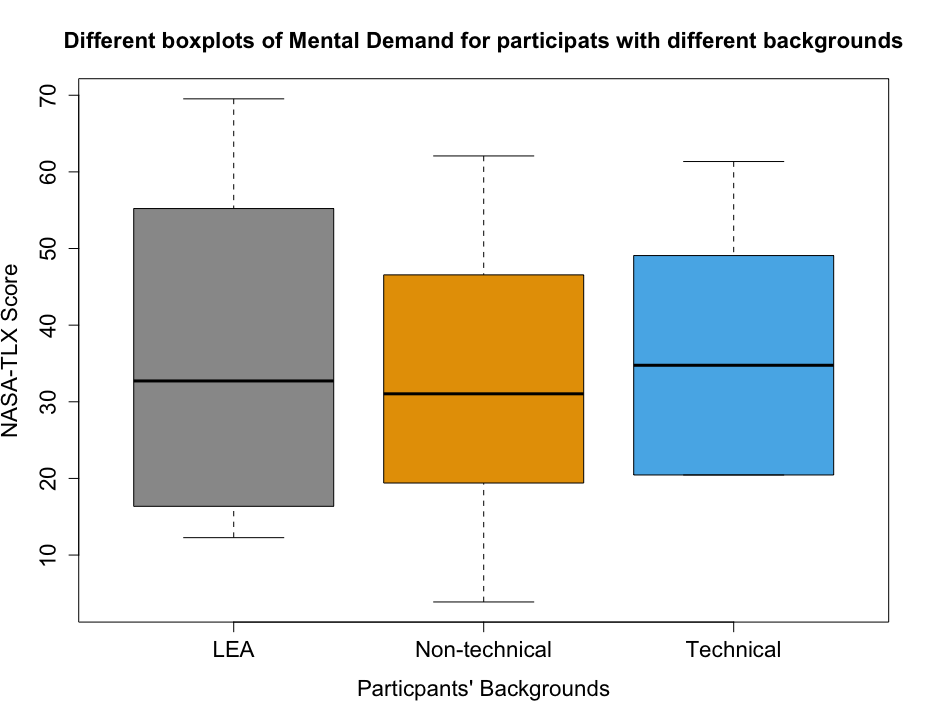}
        \caption{}
    \end{subfigure} 
    \begin{subfigure}{0.35\textwidth}
        \includegraphics[width=\textwidth]{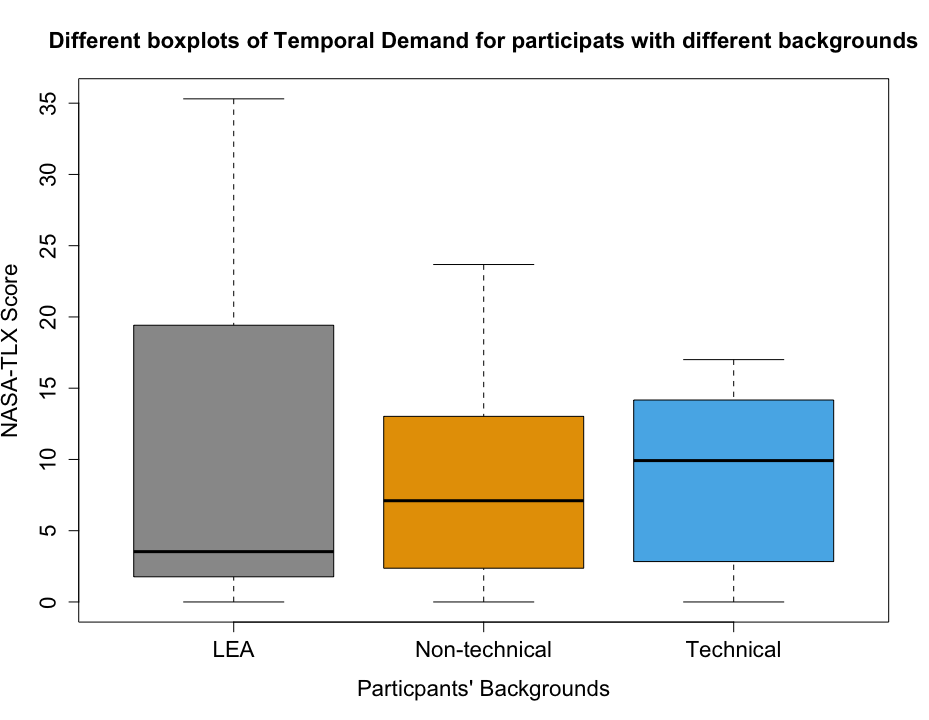}
        \caption{}
    \end{subfigure}
    \begin{subfigure}{0.35\textwidth}
        \includegraphics[width=\textwidth]{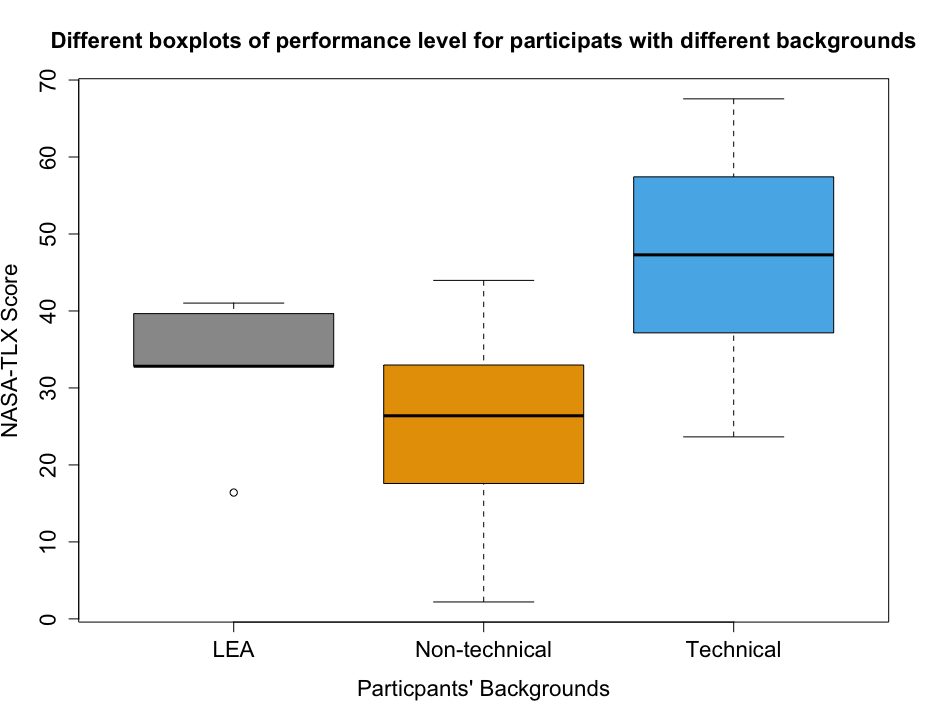}
        \caption{}
    \end{subfigure}
    \begin{subfigure}{0.35\textwidth}
        \includegraphics[width=\textwidth]{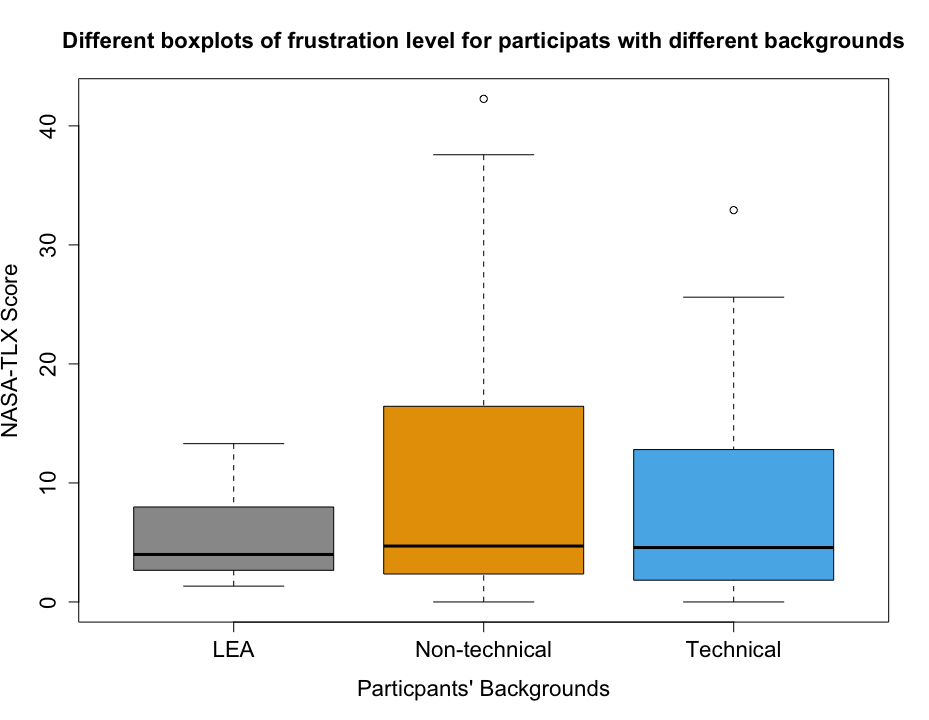}
        \caption{}
    \end{subfigure}
    \vspace{-5pt}
    \caption{\textbf{Boxplots of the weighted NASA TLX Scores of participants with different backgrounds}}
    \label{boxplots}
\end{figure}

\subsubsection{Assessing Participants' Mental Workload During the Interaction With Crime Mapping.}
We wanted to assess participants’ level of engagement while interacting with the application and their mental workload as a result of this interaction. To accomplish this, we conducted the NASA-TLX survey and assessed participants on four factors (i.e., mental demand, temporal demand, performance, and frustration level). To assess the differences in scores across participants and their significance, we conducted quantitative analysis, the details of which are provided in \hyperref[thirdappendix]{Appendix C}.

In sum, we found that all three groups of participants found the task to be mentally demanding as all three groups scored high on mental demand (MD). In addition, the average weighted MD scores for LEAs were higher than those of community members and technical participants. We also observed this during our observation where LEAs were deeply engaged in interpreting the map and connecting the hotspots to their street-level knowledge. We observed a similar pattern regarding temporal demand (TD) (i.e., stress experienced due to time constraints) where LEAs spent a significant amount of time interpreting the map, however, the community members and technical participants spent more time generating new maps, investigating how the application worked, and connecting their contextual knowledge of different neighborhoods to hotspots they saw on the map.  

The third parameter we considered was performance (PF) which determines the success of the users in carrying out their duties and how satisfied they were with the results of their work. Both the LEAs and technical participants scored high on this parameter which depicts that these participants were confident about their results. On the other hand, community members’ scores were in the medium range implying that they were uncertain about their results. As depicted in the previous section, both the LEAs and technical participants had some experience working with crime maps or geospatial heatmaps. However, technical participants interpreted maps based on their understanding of heatmaps whereas LEAs interpreted maps based on their street-level domain knowledge.

The fourth parameter we considered was the frustration (FT) level which determines whether the task was frustrating or mentally demeaning where the users were left feeling insecure, desperate, offended, or disturbed by the task. Both the LEAs and technical participants scored low on this parameter indicating that they did not find the task to be frustrating. Community members, on the other hand, scored in the medium range on this parameter. During the interview, we learned that community members found it to be frustrating to continually change parameters to go back and forth between maps of different complexities and would have preferred to be able to compare them alongside each other.

Taken together, the results for LEAs are surprising. The NASA-TLX results show that LEAs were deeply invested in interpreting the maps presented to them (i.e., they scored high on mental and temporal demand), considered that they performed well when interpreting the hotspots and placing squad cars (i.e., scored high on performance) and did not find the task to be frustrating (i.e., scored low on frustration). However, as noted previously, none of the LEAs changed parameters to generate different maps. Here, LEAs were more interested in posthoc analysis and interpreting the information provided to them instead of generating new maps.


\subsection{Participants' Needs Regarding the Ethical Design \& Use of AI Crime Mapping (RQ3)}
\subsubsection{The Need for Oversight and Creating the Right Incentives.}
Several community members and technical participants found utility in algorithmic crime mapping as a public resource, however, questioned the underlying motivation of police departments themselves. Much of our conversations in this regard centered around ongoing policing practices where they questioned the trustworthiness of police reports by drawing on cases of police brutality where the police reports did not match the video evidence shared by bystanders on social media. For instance, participant C21 shared:

\begin{quote}
    \small{"From my personal experience, it seems that in black neighborhoods police are stopping more drivers per month per hour for all kinds of things, and for things that they don't stop white drivers in the suburbs for. There are penalties for more and more people for using illegal drugs in the city, but people living in the suburbs use the same drug. But they get off with a warning and they think they are somehow different and too good for prosecution and they get some kind of a better deal. So there is a wide widespread perception in society, that law enforcement is not really fair. And on one hand, I would like to believe 'no, it really is fair'. But the evidence is coming up more and more in recent years. Like with the development of the cell phone, people take photographs and take videos and it doesn't match the police reports. Before we had this technology, and some person who was arrested would say that the police beat him up and we would hear the police officer say that `he was resisting arrest'." - C21, Community Member}
\end{quote}

\vspace{-0.1cm}
This concern was shared by several community members who believed that such technologies could lead to racial profiling and amplify racial disparities. Participants shared examples of cases where the defendants were wrongfully convicted by association and forced into plea deals. Here, technical participants also brought into question the objectivity of the data analysis process itself and the need for public transparency and accountability mechanisms that keep such technologies in check. For instance, participant T5 shared the following when discussing the role of a data analyst working in a crime analysis lab: 

\begin{quote}
    \small{"Even if this guy went through lots of schooling and got like three PhDs and is studying crime analyses, algorithms, maybe not three, let's just say one. This guy has one PhD in crime analysis algorithms. Would you kind of believe him? If he's saying that it's `efficient and fair'. I would take it with a grain of salt. Because if nobody is challenging him on anything, this guy could just be so narrowminded in any one particular mindset that he could be completely oblivious to other aspects. -T5, Technical Participant}
\end{quote}

\vspace{-0.1cm}
This concern was also echoed by other technical participants who shared examples of how data can be tweaked and cherry-picked and drew attention to the subjective nature of data science practice, an ongoing concern that has been raised in prior work \cite{mothilal2024towards, das2024colonial, passi2017data, passi2020making}. Finally, as discussed in Section 4.1, participants shared examples of differences in policing practices in urban and suburban neighborhoods, examples of speed traps, and people in low-income neighborhoods being set up for petty crimes to draw attention to policing practices that unethically target vulnerable communities instead of protecting and serving them. Here, community members sought accountability from police departments regarding street-level policing practices, whereas, technical participants wanted more transparency and accountability regarding data science practices that drive algorithmic crime mapping and data-driven policing.

\vspace{-0.1cm}
\subsubsection{Tensions Around the Ethical Basis of Data Collection}
Several community members discussed the need for trustworthiness and accountability regarding data collection practices and discussed instances of falsehoods in police reports, differences in data that is available about urban versus suburban communities, and policing practices designed to target people (e.g., speed traps). Here, some participants also recognized that several government agencies collect data about citizens while conducting their daily operations but wanted to ensure that it is collected lawfully and with some oversight from the judicial system. For instance, a participant shared: 

\begin{quote}
    \small{"I always tell them [government officials] that they need to get whatever documentation [legal paperwork], whatever court order it is, to get my information. I need the proof that they need it. I need to know what they're going to be doing with it. I'd want to know, pretty much as much as I possibly can about it. I want to know who's entering it? And want to know where it's being entered? How can anyone access it? Because in that case, you have to consider there are also hackers. So I mean, I’d want to know how secure the databases that they have are. Because that's my personal information. And if they're putting it out there, then someone is going to be able to grab it if they see fit. Not that I'm a target. But privacy." -C12, Community Member}
\end{quote}

\vspace{-0.1cm}
This concern was also echoed by LEAs who shared that it was necessary to collect data legally and ethically otherwise it could not be used for investigations and would not be admissible in court during legal proceedings. An LEA shared:

\begin{quote}
    \small{"A few years ago, they were having problems with mobs of **-year-olds running around destroying property. They hit [got access to] the city’s cell phone data to kind of.. more or less.. hack data and find out whoever these kids were... it seemed pretty effective, but of course, there needs to be, at some point, there needs to be some kind of a judge involved if they’re going to be tapping into people’s personal data like that." - L4, LEA}
\end{quote}

\vspace{-0.1cm}
Here, the LEA provides an example of the police department getting access to people’s cell phone data in a specific neighborhood, however, states that they must make their case before a judge and get a court order approved before receiving access. LEAs also shared that their real-time crime mapping technology was based on service calls and not police reports and that people’s biases played a significant role regarding the calls they receive and investigate: 

\begin{quote}
    \small{"A big part of it for me is how was the call actually generated? Did somebody call in and go, “hey, there’s this dude over here who’s suspicious.” And then when you ask them, why they’re suspicious, they can’t give you a reason... which usually means ‘walking while black’, which for PD means you still have to send officers. But then it makes the department look racist, because we’re required to check it out, because people are calling in and telling us they’re suspicious." - L3, Law Enforcement Agent}
\end{quote}

\vspace{-0.1cm}
Several LEAs recognized the concerns surrounding racial profiling but drew attention to the nature of their work where they were required to investigate service calls as well as patrol the zones that were assigned to them. Similarly, several LEAs shared the nuances of their street-level labor that we discuss next.

\vspace{-0.1cm}
\subsubsection{Nuances of Policing Practices}
LEAs drew attention to several nuances of street-level practices that needed to be considered and their impact on decision-making. Participants shared that technologies such as algorithmic crime maps were useful in assessing if/when there was increased criminal activity in specific neighborhoods but did not help answer why this might be the case. The crime map may help create a temporary specialized patrol zone but these cases still need to be investigated. LEAs needed to understand the borders between districts, relationships between different types of crimes, and context-specific information that they receive from people on the grounds. For instance, an LEA shared - 

\begin{quote}
    \small{"It influences my decision-making process as to where I’m going to spend more of my time researching crimes in this area versus another... let’s say [anon] Street, for example. Why are we seeing such a high increase in crime from gangs? Is it because it’s a border for two districts? Is it because we’re not allocating enough resources to that area and it’s allowing crime to thrive? What’s the root cause?" -L5, Law Enforcement Agent}
\end{quote}

\vspace{-0.1cm}
Similarly, LEAs also shared that they needed trained investigators who could interview community members and get consistent information about what was happening in the neighborhood that led to increased criminal activity. That is, LEAs needed to understand the changing street-level dynamics to assess the root cause of certain crimes. LEAs also mentioned that a better understanding of street-level practices was essential for any systems use. They were frustrated that oftentimes these systems did not align with how work was conducted. For instance, an LEA shared -

\begin{quote}
    \small{"I never knew the algorithm that was used - that was never shared. There were a couple of occasions where me and my team would be given a packet of information about a specific crime that was occurring in our area of responsibility, and sometimes the predictive information that was given to us wasn’t consistent with what we knew as law enforcement officers just being in these neighborhoods." -L5, Law Enforcement Agent}
\end{quote}

\vspace{-0.1cm}
Here, LEAs also shared that it was essential for any data analyst to understand the nuances of street-level labor because there is a significant amount of discretionary work that LEAs undertake. In the quote below, an LEA shares that they would have significant concerns about data-driven policing -   

\begin{quote}
    \small{"If they [data analysts] didn’t have any background, in working with a police department, they have just the degree in… you know… computer science or data science or whatever, because, I mean, you can learn a lot of your crime stuff on the job. ...However, if someone with zero experience in interacting with people and police departments came in and said, “Hey, we’re going to do things by the numbers now”, I’d be very much concerned." - L2, Law Enforcement Agent}
\end{quote}

\vspace{-0.1cm}
Here, it is interesting to note that both the technical participants and LEAs drew attention to the nature of data science practice. Technical participants discussed the elastic and subjective choices that data scientists make during analysis, whereas, LEAs discussed the need for data scientists to understand the nature of street-level discretionary work that they undertake. These themes have been observed across different contexts where domain experts have drawn attention to nuances of their professional practice that cannot be accounted for by quantitative data, and are generally unobservable to people outside the profession but significantly impact the decision-making process \cite{kawakami2022improving, watkins_art, spektor2023designing}.

\section{Discussion}
In this section, we first discuss how engaging with community members at the earliest stages of the AI deliberation process can help inform machine learning (ML) problem formulation and help proactively identify ethical issues before the systems are developed and deployed. Next, we discuss how AI applications used as boundary objects can help facilitate deeper and nuanced conversations with community members, and finally, we discuss some implications for human-AI decision-making and provide design guidelines for AI systems that must serve the public interest.

\vspace{-0.2cm}
\subsection{Eliciting Feedback from Community Members to Inform AI Problem Formulation}
Early deliberations regarding the development and use of AI systems often omit impacted stakeholders because of their lack of technical or data science knowledge which supposedly limits them from providing meaningful feedback that can inform systems design \cite{ kuo2023understanding, delgado2023participatory}. However, as our results highlight, \textbf{community members, irrespective of technical knowledge, are able to provide detailed feedback regarding whether the technical formulation of a problem is even socially relevant, draw attention to systemic and structural issues that might be invisible to AI developers, and help uncover the downstream impact of such systems on their communities.} That is, impacted stakeholders have a critical role to play at the earliest stages of machine learning problem formulation where data scientists translate broad and nondescript objectives into tractable problems that machine learning can answer \cite{passi2019problem}. Our results show that community members drew upon their lived experiences and questioned the motivation behind the application, such as challenging the police department's claim of `efficient resource allocation,' which they argued may unintentionally exacerbate racial profiling in low-income neighborhoods. Participants also question whether increased police presence truly leads to safer neighborhoods, further suggesting a need to reconsider the predominant focus on quality-of-life crimes.

Here, community members also shared that there needed to be transparency regarding the use of such tools by government agencies but also that they needed to operate in the public interest; i.e. - improving the safety of citizens through fair policing practices that do not indiscriminately target and racially profile members of the community. Community members brought into question the street-level practices of law enforcement agents through which data is collected about low-income and minority communities. Similarly, technical participants, who are also community members, shared that such policing practices might perpetuate a vicious cycle by consistently collecting crime data in low-income neighborhoods. This cycle, or feedback loop, could result in increased policing and continuous data collection, a concern also raised in prior work \cite{veale2018fairness}. In line with recent work in the public sector \cite{kuo2023understanding, stapleton2022has, zhang2023deliberating}, our results provide further evidence that impacted community members are indeed indispensable to the AI deliberation process and can help avoid major pitfalls in AI systems. That is, \textbf{involving impacted community members at the onset of the AI deliberation process can help proactively identify critical fairness, accountability, transparency, and ethical (FATE) issues before AI systems are developed and deployed; a significant ongoing concern for Responsible AI (RAI) practices.} As highlighted by Kawakami et al. \cite{kawakamistudying}, this concern was also raised by public sector workers who believed that collaborating with community members could help address "fundamental and root problems" that are missed otherwise.

Here, the application acted as a boundary object \cite{star1989structure} and helped facilitate detailed conversations with community members. The application helped scaffold nuanced and context-specific discussions with community members about policing practices in their city, and the use of AI systems by police departments, and helped articulate design guidelines for developing ethical AI systems that are centered in the needs of the impacted community.

\vspace{-0.2cm}
\subsection{AI Applications as Boundary Objects and Educational Tools}

Through our observations of lab activities conducted with community members, we learned that algorithmic crime mapping served as an educational tool that taught them several key aspects of algorithmic decision-making. First, the application focused on familiar city neighborhoods which helped establish a degree of comfort in the lab activity such that community members were not overwhelmed and were able to freely deliberate. Second, by engaging with parameters, community members understood that the underlying algorithm was simply generating estimates (i.e., hotspots) and not some objective truth. Their choice of parameters (i.e., size, smoothness, and distance between hotspots) impacted how the results were generated on the map. Even though we had given the participants a brief overview of each parameter, they developed a more nuanced understanding by experimenting with them (e.g., the degree of numeric change causing the degree of visual change on the map). Participants also found the results to be more interpretable because they were able to ground these empirical findings within their local and contextual knowledge of different neighborhoods. 

Several community members and technical experts unexpectedly sought to compare different maps simultaneously, causing frustration for some participants. However, a few quickly addressed the issue by using the `snipping tool' to capture and preserve the current map before generating a new one. As participants engaged with the maps and understood the underlying data, they started to raise concerns regarding issues such as speed traps, racial profiling in low-income areas, and policing disparities between suburban and inner-city regions. Here, the application acted as a boundary object that facilitated participants in brainstorming issues and initiating discussions with the researchers.  However, some participants found the tool useful if operated in the public interest, envisioning it as a resource to learn about different neighborhoods. For example, a participant suggested using the tool to figure out if a neighborhood was prone to car thefts and subsequently decide whether to purchase garage parking. \textbf{By engaging and deliberating with the application, participants were able to develop an informed mistrust of the AI tool, i.e. - they were able to think of use cases where the AI can be used responsibly and inform decision-making but also articulated scenarios where the application could do harm (e.g., racial profiling)}. In sum, this study provides further evidence that integrating user control into intuitive ADS that leverages local and contextual knowledge can effectively anchor AI concepts and educate community members. This is particularly significant given the expanding research focus on public interest technology \cite{stapleton2022has, aragon2020civic, mccord2023beyond}, community-driven AI \cite{hadley2022covid, kuo2023understanding, stapleton2023seeing}, and expanding participation in AI design \cite{delgado2023participatory, zhang2010detecting, kuo2023understanding}  within the SIGCHI community.

On the other hand, our observations of domain experts (i.e., LEAs) revealed a more complicated story. LEAs demonstrated a tendency to anchor their street-level knowledge to specific landmarks and neighborhoods in the city, persistently sticking to the initial parameters despite the researchers urging them to explore different maps. This indicated anchoring bias, a cognitive bias where individuals rely too heavily on the first piece of information they encounter (the "anchor"). Anchoring bias has been observed in domain experts across various disciplines \cite{furnham2011literature, azzopardi2021cognitive, saposnik2016cognitive}. However, as noted in Section 4.3.3, LEAs shared that the application lacked essential features for deriving meaningful information. As a result, they engaged with it at a higher level, using it to validate street-level knowledge and generate evidence for requesting specialized patrol zones. Our findings provide further evidence that AI systems that are meant to augment workers' \textit{social decision-making} \cite{kawakami2022care} must seek to complement their decision-making abilities and overcome their limitations \cite{kawakami2023training, guerdan2023ground}. That is, \textbf{worker-centered design principles are fundamental for laying the groundwork for human-AI decision-making and evaluating whether human+AI decisions lead to improved outcomes.} We delve deeper into this perspective in the next section.

\vspace{-0.2cm}
\subsection{Implications for Human-AI Decision-Making}
We learned that domain expertise played a significant role in determining LEAs’ interactions with the application. LEA used their domain expertise at all three levels of engagement: understanding the system, interacting with the system, and interpreting the results; highlighting that their decision-making was a function of their domain (i.e., theoretical) knowledge where the application (i.e., empirical knowledge) was mostly used to anchor and validate pre-existing information. That is, empirical information did not augment the practitioners’ knowledge by introducing factors they would not have considered otherwise. This was evident from the LEAs choosing not to generate new maps as well as the results of the NASA-TLX survey that showed that they were mentally engaged but in doing so relied too heavily on the initial information presented to them which in turn influenced their subsequent decisions. This further suggests that there is a need for LEAs to unlearn their current ADS practices and be trained to engage in \textit{critical use} \cite{kawakami2023training} - engaging in appropriate reliance by situating AI predictions against their contextual and domain knowledge.  

\textbf{We also learned that target-construct mismatch \cite{kawakami2023training, guerdan2023ground} can exist even in applications as intuitive as algorithmic crime mapping.} Predicting hotspots across the city is supposed to help LEAs distribute and allocate resources more efficiently to deter crime (i.e., the target variable), however, we learned that LEAs were more interested in finding critical hyperlocal regions such as key intersections to create specialized patrol zones (i.e., construct of interest) where they could direct resources and also ensure the availability of backup for patrol cars. Here, as highlighted by Kawakami et al. \cite{kawakami2023training}, effective AI use depends on the practitioners’ ability to account for such misalignments.

Here, practitioners can find themselves translating information from these two different sources which can often increase uncertainty and lead to unreliable decisions \cite{kawakami2022improving, mcconvey2023human, saxena2022train}. \textbf{Systems that are not designed to complement the workers’ expertise (i.e., grounded in worker-centered design \cite{fox2020worker, collective2022human}) can further complicate human-AI decision-making and make it difficult to assess its efficacy.} This argument may seem intuitive but needs to be clearly formulated. Domain experts are being asked to think in counter-intuitive ways and adopt empirical models that introduce data-driven insights to decision-making. However, without proper heuristic guidelines regarding how to incorporate these insights within decision-making such that human+AI decisions are an improvement over human decisions, such empirical models can either become a source of frustration for decision-makers (see for e.g., \cite{cheng2022child, saxena2023rethinking, kawakami2022improving, watkins_art}) or otherwise, as depicted in this case study, can be a source of anchoring bias. This was also evident from the LEA’s response who claimed that they were only looking for a satisfactory outcome from the system and not looking to derive new spatiotemporal information to augment their decisions.

Therefore, to be able to evaluate the efficacy of human-AI decision-making, it is first necessary to ensure that the AI complements the workers’ expertise and augments their decision-making in specific ways. For instance, LEAs shared that it would be meaningful to draw out the boundaries between districts (i.e., insert their hyperlocal contextual knowledge onto the map), look at the relationship between different types of crimes to assess any spatiotemporal relationships, or incorporate information derived from on-the-ground investigations. That is, \textbf{worker-centered design of AI applications can help us better approach the evaluation of human+AI decisions by ensuring that AI complements workers’ expertise and augments their decisions in measurable ways.} This common theme regarding human-AI decision-making has been observed across contexts (e.g., child welfare \cite{kawakami2022improving, saxena2023algorithmic}, nursing \cite{watkins_art}, hospitality \cite{spektor2023designing}) where workers drew attention to the nuances of their labor that the AI did not account for and which inadvertently augmented uncertainty and left them performing added labor to repair the disruption in processes caused by the AI system.

Here, ADS such as algorithmic crime mapping could be redesigned to be worker-centered by incorporating LEAs' feedback. For instance, contextualizing empirical data points with qualitative narratives from police investigations where explanations (i.e., a summary of key points) are generated alongside hotspots to help LEAs assess if new patterns of criminal activity are emerging in neighborhoods and inform collaborative decision-making (See Ehsan et al. \cite{ehsan2023charting} for important practical considerations for designing for explanations where the nature of practice is highly collaborative).

\vspace{-0.2cm}
\subsection{Design Guidelines for Algorithmic Crime Mapping that Serves the Public Interest}

Our study of a prominent ADS used in the criminal justice system found some comprehensive design principles for creating AI systems for public sector settings. However, as researchers, we must contend with the fact that most of the algorithmic tools used by law enforcement agents in the United States are privately developed with little room for comments from the public. In addition, considering the data being used by such technologies and their disparate use in urban areas for the policing of racial and low-income minorities, we believe that such technologies can lead to significant representational, interpersonal, and social algorithmic harms \cite{shelby2023sociotechnical}. Therefore, it was essential to explore the community members' perspectives and gather their feedback. Community members drew upon their lived experiences, the technical participants used their data science knowledge, and the LEAs leveraged their domain expertise as they deliberated while interacting with the application. Below, we provide some design guidelines --

\begin{itemize}
    \item Community members, drawing on their lived experiences, highlighted policing practices that led to racial profiling. They sought increased transparency regarding the use of AI systems and questioned the core objective of AI crime mapping as formulated by developers and police departments.
    
    \item Technical participants discussed the elastic nature of data science practice and recommended a more collaborative process for decision-making to ensure accountability and transparency around the practices of crime analysts. They also drew attention to the harmful, data-driven feedback loop created by algorithmic crime mapping.
    
    \item LEAs suggested several ways in which the AI could complement their domain expertise. They wanted to be able to draw boundaries for different districts, assess spatiotemporal relationships between different crimes, as well as see information from on-the-ground investigations.
    
    \item Worker-centered design of AI systems, as specified in the point above, can further make the system more transparent to workers by supporting AI explanations \cite{ehsan2021expanding} that are grounded in the nature of practice.
    
    \item It is important to recognize that a target-construct mismatch \cite{kawakami2023training} is likely to exist, assess its impact on predicted outcomes, and whether the predicted target still aligns with the practitioners' decision-making goals.
    
    \item In response to algorithmic unfairness embedded in sociotechnical systems, significant attention has been paid to the biases embedded in datasets (e.g., disproportionate impact on some social groups). By putting the practitioners' contextual knowledge into action, it is necessary to assess any process-oriented harms (i.e., harm to the decision-making process itself) that may occur \cite{saxena2023algorithmic}. 
    
    \item AI systems need to incorporate contestability with key stakeholders. Domain experts need to be able to challenge decisions made at lower levels of implementation (such as resource allocation decisions made by the CJS), and community members need to be able to challenge decisions made at higher levels of implementation (i.e., the driving motivation and consequent problem formulation).
\end{itemize}

\section{Limitations}
Our study has a limitation in terms of participant distribution across different categories. Our primary aim was to conduct the experiment with real community stakeholders, and thus we put considerable effort into recruiting individuals from a wide array of professions, with a specific focus on those with law enforcement experience. While the number of participants in this category may be lower than in the other two groups, we successfully enlisted officers from various ranks and departments, ensuring a comprehensive perspective. Furthermore, our recruitment strategy did not confine us to specific city regions, such as university areas. Instead, we adopted an inclusive approach by disseminating recruitment information throughout the city, allowing us to capture voices from diverse genders, races, ages, and income levels. The experiment itself was conducted in a midwestern, mid-sized metropolis and the results may not represent community stakeholders' perspectives in other regions of the country, especially major cities. However, we deliberately selected this city for our study because it provided us with insights from a culturally diverse population, which better reflects grassroots sentiments. This choice allowed us to gather a broad range of stakeholder perspectives, in contrast to larger cities where obtaining a representative sample of such diverse opinions can be more challenging.

\section{Conclusion}
In this study, we conducted a mixed-methods investigation in a midwestern, mid-sized city in the United States. The goal of this study was to assess how community stakeholders from different backgrounds interacted with and interpreted algorithmic crime mapping. To achieve this, we designed an in-lab activity where participants were asked to interact with the tool, and subsequently, identify hotspots on the map. Additionally, we sought to gain insights into their perspectives through semi-structured interviews conducted immediately after the activity. We learned that community members found utility in algorithmic crime mapping but were concerned about the incentive structures within police departments themselves that often racially profile and target low-income communities. Technical participants, on the other hand, drew attention to the elastic and subjective nature of data science practice and the need for a more collaborative and interdisciplinary process for decision-making. We learned that the LEAs' interaction and interpretation of this application was significantly informed by their domain knowledge and they often used it as a means of validating their pre-existing street-level knowledge instead of trying to derive new information from it. We further show that AI systems need to be designed using worker-centered design principles if they are to augment human decisions such that human+AI decisions are an improvement over human-only and AI-only decisions.


\begin{acks}
This research was supported by the National Science Foundation grant (CRII-1850517) and the Natural Sciences and Engineering Research Council of Canada (NSERC) grant (RGPIN-2022-04570). Any opinions, findings, conclusions, or recommendations expressed in this material are those of the authors and do not necessarily reflect the views of our sponsors or community partners.
\end{acks}


\bibliographystyle{ACM-Reference-Format}
\bibliography{MAIN}

\appendix

\vspace{-0.1cm}
\section{ESTIMATING THE NUMBER OF HOTSPOTS}\label{firstappendix}
\begin{figure*}[t!]
\includegraphics[width=0.9\textwidth]{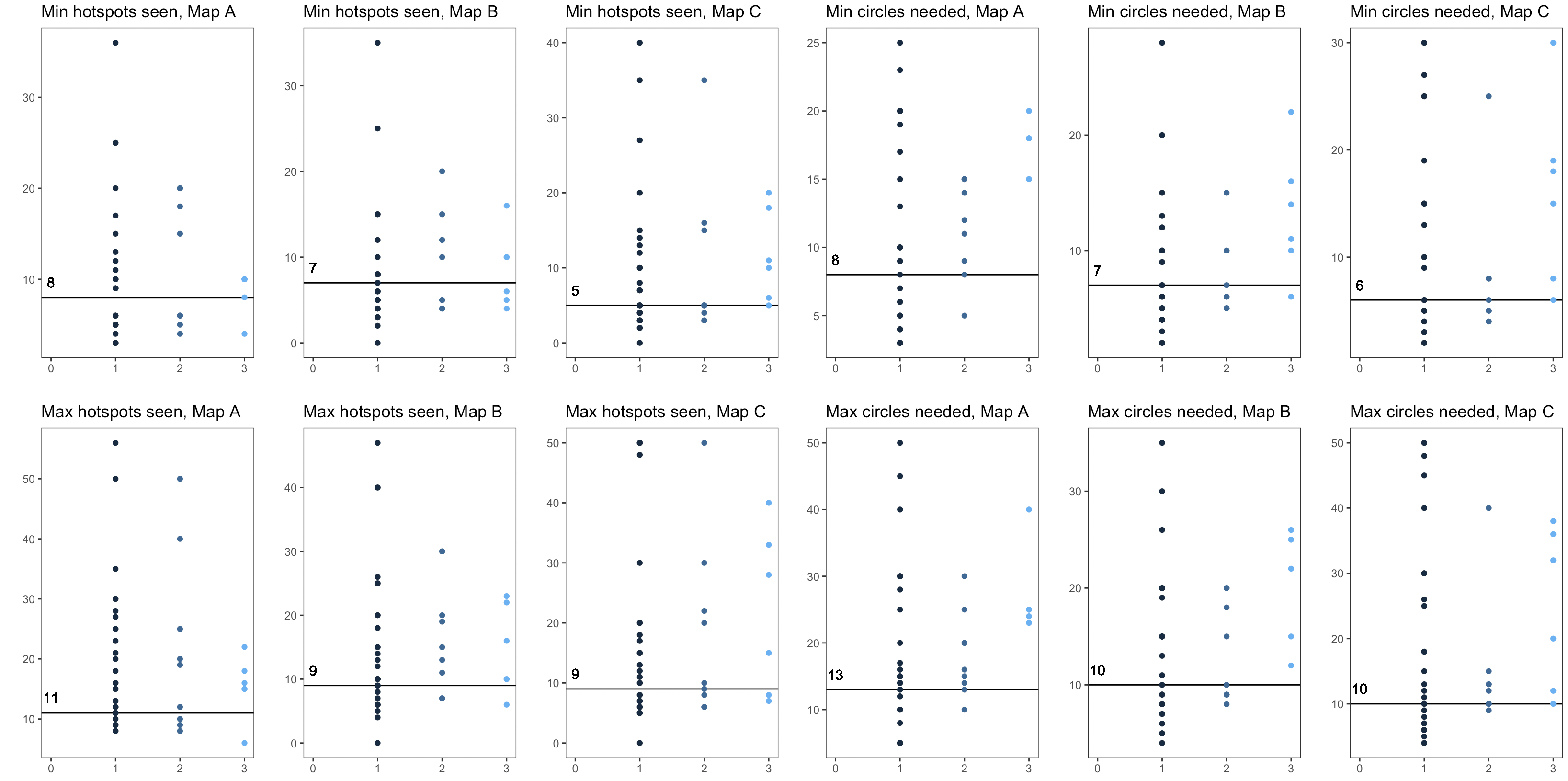}
\vspace{-5pt}
\caption{{\textbf{Estimates of hotspots (left) and circles (right) seen across groups and maps compared to actual values}}}
\label{descHotspots}
\vspace{0.5cm}
\end{figure*}

Fig \ref{descHotspots} shows the distributions of minimum and maximum hotspots grouped by background. For our statistical analysis, we compared user-selected values to actual values for each map type (the solid line represents actual values). We conducted this analysis to ascertain whether people from different backgrounds tend to over or under-estimate hotspots on the crime maps.  

On average, members of Group 1 (i.e., community members) estimated minimum hotspots to be an average of 8.82, and maximum hotspots to be 13.61 more than the actual value. This is significantly lower than the averages of Groups 2 (i.e., technical participants) and Group 3 (i.e., LEAs), who estimated minimum hotspots to be greater by 15.6 and 13.43 respectively, and maximum hotspots by 21.17 and 18.29 greater than the actual value. Thus, our first takeaway from these results is that, on average, all three groups of participants overestimated hotspots from maps of varying complexities but rather counter-intuitively, community members overestimated hotspots to a much lower degree than LEAs. This has several important implications for the interpretation of algorithmic crime mapping that we discuss in Section 4.2.1.  


\vspace{-0.1cm}
\section{Interaction With Parameters For Creating New Maps}\label{secondappendix}

\begin{figure*}[t!]
\includegraphics[width=1\textwidth]{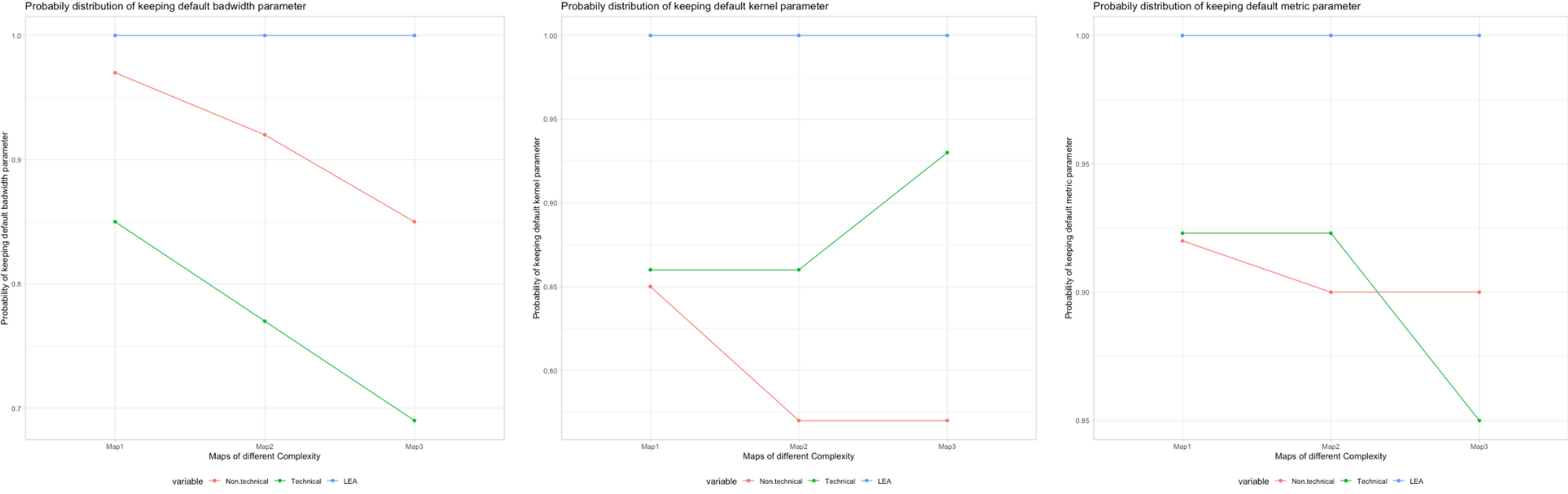}
\vspace{-5pt}
\caption{{\bf Interaction plot of keeping default kernel settings (left), keeping default bandwidth settings (middle), default metric settings (right) across different backgrounds. Here, red points represent non-technical/community members (Group 1); green points represent technical members (Group 2); blue points represent LEAs (Group 3).}}
\label{interactiondefault}
\end{figure*}

Fig \ref{interactiondefault} shows how different user groups changed the default parameters: solid line for technical participants, dotted line for LEAs, and hyphened line for community members. In transitioning from maps A to B, LEAs maintained a constant likelihood of retaining the default kernel parameter, while community members became less inclined to do so. As complexity increased from maps B to C, technical participants were more likely to keep the default kernel parameter, whereas community members maintained a constant probability. As maps grow in complexity, LEAs consistently showed a probability of not changing the bandwidth parameter. Meanwhile, both technical participants and community members exhibited a decreasing likelihood of keeping the default bandwidth parameter. With increasing complexity, both groups were more likely to deviate from the default. Finally, LEAs maintained a constant probability of not altering the default distance metric parameter. Our qualitative analysis (see Section 4.2.3) suggests varied motivations for parameter changes among users of different backgrounds. For technical participants, there is a consistent probability of keeping default parameters from A to B but a decreased likelihood of retaining the default metric parameter from B to C. As map complexity rises from medium to hard, technical participants showed a greater inclination to change the default metric parameter, reflecting a mix of familiarity and curiosity.


\section{Assessing mental workload of tasks}\label{thirdappendix}

We used the NASA-TLX survey to gauge participants' mental workload while using the application. Focusing on factors related to psychological demand, mental demand, temporal demand, performance, and frustration level, we analyzed scores for participants across the three groups. Descriptive statistics, shown in Fig \ref{boxplots}, highlight the score variations. We assessed mental workload using the first parameter, mental demand (MD), which measures mental and perceptual activities. The weighted scores in Figure 8 indicate a somewhat high MD (30-40 out of 100) for all three groups, suggesting a moderately challenging task. LEA participants had higher average MD scores than community members and technical participants. However, the data's higher standard deviation and lower median for LEA suggest more variability, with some finding the task to be mentally challenging. This reflects LEA's perception of the task as being important regarding understanding the map to allocate patrol cars effectively. 

We considered temporal demand (TD) as the second parameter, assessing time pressure during the task. Similar patterns to mental demand (MD) in descriptive statistics are also observed in TD. The correlation between mental demand (MD) and temporal demand (TD) suggests that the patterns and reasoning in mental demand scores apply here to the observations of temporal demand.

The third parameter we considered is performance (PF) which gauges workers' success and satisfaction. In Figure 8, community members have medium PF scores (10-29), while technical participants and LEAs show somewhat high scores (30-49) \cite{NASA}. This suggests greater confidence in results for technical participants and LEAs compared to community members. We explain in Sections 4.1.1 and 4.1.2 that LEAs and technical participants are more familiar with such ADS than community members. Despite technical participants having data science expertise (average PF score of 45.6), LEAs, due to prior experience working with algorithmic crime maps, exhibit similar confidence levels in their outputs.

The fourth parameter we considered is frustration level (FT), which assesses whether the job is frustrating or mentally demeaning such that the participant is left feeling insecure, desperate, offended, or disturbed when doing the task. The average FT is medium (10-29) for community members and low (0-9) for technical participants and LEAs. As we highlight in the qualitative results, community members were more frustrated because they wanted to generate and compare different maps. Notably, LEAs were less frustrated than technical participants, likely because they didn't generate new maps, relying on the first map and connecting it with their street-level domain knowledge. Through this analysis, we can say that the mental workload regarding the task is not very high which implies it is feasible to use such AI applications as boundary objects that can facilitate deeper conversations about algorithm design and use with a variety of stakeholders.

\end{document}